\documentclass[fleqn,11pt]{article}

\usepackage{amsmath,amssymb,amsfonts,graphicx}

\usepackage{amsfonts,amssymb,cite}
\usepackage{graphicx}

\def\onecol{\onecolumn \mathindent 2em}

\def\noi{\noindent}

\newcommand{\Title}[1]{\noi {{\Large\bf #1}}\\[1ex]}

\def\Aunames#1{\noi{\bf #1}}
\def\au#1{${}^{#1}$}
\def\Addresses#1{\medskip\noi \protect
	\begin{description}\itemsep -3pt {\it #1} \end{description}}
\def\adr#1#2{\item[${}^{#1}$]{\it #2}}

\newcommand{\Abstract}[1]{\vskip 2mm \begin{center}
        \parbox{16.4cm}{\small\noi #1} \end{center}\medskip}

\def\email#1#2{\footnotetext[#1]{e-mail: #2}\addtocounter{footnote}{1}}

\def\nqq{\hspace*{-2em}}
\def\nhq{\hspace*{-0.5em}}

\def\qq{\qquad}

\def\ten#1{\mbox{$\times 10^{#1}$}}
\usepackage{color}

\def\Acknow#1{\subsection*{Acknowledgments} #1}

\def\Jl#1#2{#1 {\bf #2},\ }

\def\ApJ#1 {\Jl{Astroph. J.}{#1}}
\def\CQG#1 {\Jl{Class. Quantum Grav.}{#1}}
\def\DAN#1 {\Jl{Dokl. AN SSSR}{#1}}
\def\GC#1 {\Jl{Grav. Cosmol.}{#1}}
\def\GRG#1 {\Jl{Gen. Rel. Grav.}{#1}}
\def\IJMPD#1 {\Jl{Int. J. Mod. Phys. D}{#1}}
\def\JETF#1 {\Jl{Zh. Eksp. Teor. Fiz.}{#1}}
\def\JETP#1 {\Jl{Sov. Phys. JETP}{#1}}
\def\JHEP#1 {\Jl{JHEP}{#1}}
\def\JMP#1 {\Jl{J. Math. Phys.}{#1}}
\def\NPB#1 {\Jl{Nucl. Phys. B}{#1}}
\def\NP#1 {\Jl{Nucl. Phys.}{#1}}
\def\PLA#1 {\Jl{Phys. Lett. A}{#1}}
\def\PLB#1 {\Jl{Phys. Lett. B}{#1}}
\def\PRD#1 {\Jl{Phys. Rev. D}{#1}}
\def\PRL#1 {\Jl{Phys. Rev. Lett.}{#1}}

\def\al{&\nhq}
\def\lal{&&\nqq {}}
\def\eq{Eq.\,}
\def\eqs{Eqs.\,}
\def\beq{\begin{equation}}
\def\eeq{\end{equation}}
\def\bear{\begin{eqnarray}}
\def\bearr{\begin{eqnarray} \lal}
\def\ear{\end{eqnarray}}
\def\earn{\nonumber \end{eqnarray}}
\def\nn{\nonumber\\ {}}

\def\nnn{\nonumber\\ \lal }
\def\nnnv{\nonumber\\[5pt] \lal }

\def\yyy{\\[5pt] \lal }
\def\eql{\al =\al}

\def\e{{\,\rm e}}
\def\d{\partial}

\def\diag{\mathop{\rm diag}\nolimits}

\def\const{{\rm const}}

\def\then{\ \Rightarrow\ }

\def\kappa{\varkappa}

\def\eqn#1{\eq\eqref{#1}}
\def\rf{\eqref}
\def\mN{_\mu^\nu}

\def\cK{{\cal K}}

\def\sph{spherically symmetric}
\def\ssph{static, spherically symmetric}

\def\wh{wormhole}
\def\whs{wormholes}
 
\def\emag{electromagnetic}

\usepackage{color}

\begin{document}
\onecol
\thispagestyle{empty}

\Title{Possible wormholes in a Friedmann universe}

\Aunames{Kirill A. Bronnikov,\au{a,b,c,1} Pavel E. Kashargin,\au{d,2} Sergey V. Sushkov\au{d,3}}

\Addresses{\small
\adr a {Center of Gravitation and Fundamental Metrology, VNIIMS, 
		Ozyornaya St. 46, Moscow 119361, Russia} 
\adr b  {Peoples' Friendship University of Russia (RUDN University), 
		6 Miklukho-Maklaya St, Moscow, 117198, Russia}
\adr c  {National Research Nuclear University ``MEPhI'', 
		Kashirskoe sh. 31, Moscow 115409, Russia}
\adr d {Institute of Physics, Kazan Federal University, 
		Kremliovskaya St. 16a, Kazan 420008, Russia}
		}

\Abstract{We study the properties of evolving wormholes able to exist in a closed Friedmann
	dust-filleduniverse and described by a particular branch of the well-known 
	Lema\^{\i}tre-Tolman-	Bondi solution to the Einstein equations and its generalization with 
	a nonzero cosmological constant and an electromagnetic field. Most of the results are 
	obtained with pure dust solutions. It is shown, in particular, that the lifetime of \wh\ throats 
	is much shorter than that of the {whole} \wh\ region in the universe (which coincides 
	with the lifetime of the universe as a whole), and that the density of matter near the 
	boundary of the \wh\ region is a few times smaller than the mean density of matter in 
	the universe. Explicit examples of \wh\ solutions and the corresponding numerical 
	estimates are presented. The traversability of the wormhole under study
	is shown by a numerical analysis of radial null geodesics. 
	}

% ==============================================
\email 1 {kb20@yandex.ru}
\email 2 {pkashargin@mail.ru}
\email 3 {sergey\_sushkov@mail.ru}

% ================
\section{Introduction}
% ================

  A wormhole is one of the types of strongly curved geometries, the one resembling a spatial 
  tunnel between either different regions of the same universe or different universes. 
  Such spatial geometries within solutions to the gravitational field equations were first discussed 
  in \cite{Flamm, EnstRos, Wheeler1, Wheeler2}, but those wormholes were not 
  traversable for subluminal particles or even photons, which were unable to travel from one 
  ``end of the tunnel'' to the other, to say nothing on the ability to return back. The first exact 
  solutions describing traversable wormholes seems to have appeared in \cite{Bronnikov1, Ellis} 
  in 1973 in general relativity (GR) with a massless phantom scalar field (a hypothetic field with 
  a wrong sign of kinetic energy) as a source. 
  An evolving version of such scalar-vacuum solutions was also found \cite{Ellis2} as well as 
  examples of higher-dimensional static \wh\ solutions \cite{Cle1, Cle2}. A large interest in 
  these objects has been raised by the paper of Morris and Thorne \cite{MorTho} (1988) 
  who showed that a static wormhole throat considered in the framework of GR requires 
  the existence of so-called ``exotic'' matter, violating the Null Energy Condition (NEC). 
  A phantom scalar field is a simple example of such matter. 
  
  By now, wormholes have been considered in different theories of gravity and in the presence of 
  different kinds of matter. Thus, in \cite{Bronnikov1} \ssph\ \wh\ solutions are presented both in 
  GR and a class of scalar-tensor theories, with or without an \emag\ field. Wormholes in the 
  Einstein-Maxwell-dilaton theory have been described in 
  \cite{kb95, Cle3, PriesleiGoulart, HyatHuang}. Other sources in GR used for \wh\ construction
  include a Chaplygin gas \cite{ChaplyginLobo}, various versions of phantom energy and 
  quintessence, in particular, those with the stress-energy tensor (SET) of a perfect fluid 
  \cite{Sushkov1, Kuhfittig2, Lobo3, Kuhfittig3, Sahoo, Kuhfittig4, kb17}. It was shown \cite{kb17} 
  that \ssph\ \whs\ with two flat or AdS asymptotic regions are impossible in GR with any source 
  possessing isotropic pressure, and, as a result, perfect-fluid \whs\ can only contain their source 
  in a bounded region of space surrounded by vacuum, with a thin shell on the boundary. 
  It should also be mentioned that many authors consider wormhole models built using thin 
  shells of exotic matter as the only (or main) source, the first of them being probably 
  \cite{Visser2, Visser3}. In  \cite{BlazquezSalcedo, kon21}, examples of static traversable 
  wormholes are given in Einstein-Dirac-Maxwell theory, being obtained without explicitly introducing
  exotic matter, which means that the Dirac spinor fields themselves exhibit exotic properties
  \cite{our_comm}.
    
  The necessity of exotic matter is a basic problem of \wh\ physics, at least in the case of 
  static configurations in GR \cite{MorTho, hoh-vis1} and a broad class of scalar-tensor 
  theories of gravity and $f(R)$ theories \cite{kb-star}. It is therefore natural that many authors 
  try to replace such matter with entities appearing in various extensions of GR, and above all 
  it concerns \ssph\ configurations. Thus, such wormhole solutions of asymptotically safe 
  gravity were recently discussed in \cite{AlencarNilton}. In brane world gravity, it has been 
  shown that the role of exotic matter may be played by the so-called tidal contribution to
  the effective SET due to the influence of the bulk \cite{kb-kim03}, with a number of particular 
  examples. 
In Ref.\,\cite{Harko:2013yb}, the most general constraints have been obtained on 
  additional terms inherent to various modified theories of gravity, including geometric 
  modifications, such that wormhole geometries could be constructed in such theories
  without exotic matter.
     
  It turns out that rotational degrees of freedom may in principle replace exotic matter for
  \wh\ construction. Thus, some examples of rotating cylindrically symmetric \wh\ models 
  without NEC violation have been built in the framework of GR \cite{cyl1,cyl2,cyl3}; the recently 
  found static solutions \cite{BlazquezSalcedo, kon21} involve Dirac fields with a spin; one 
  can also recall \wh\ solutions in the Einstein-Cartan theory \cite{gal1,gal2} containing no 
  exotic matter but a nonzero torsion. {Stationary rotating wormhole models with axial 
  symmetry in GR have also been obtained
  \cite{Matos, Sushkov2008, Sushkov2009, Kunz2014, Kunz2016}, however, the NEC is 
  still violated in these models of rotating \whs,
  }
	
  Dynamic \whs\ can also exist without NEC violation, at least in a finite time interval, 
  in configurations without static early-time or late-time asymptotic behavior. Such \wh\ 
  models in GR with cosmological-type metrics are known, being supported by \emag\ fields 
  described by some particular forms of nonlinear electrodynamics \cite{Arellano, NEDwh}. 
  A number of dynamic wormhole models 
  \cite{Kuhfittig3, Kar, Kim, Roman, SushkovKim, SushkovZhang} were obtained by adding 
  a time-dependent scale factor to an otherwise static metric; some others used the thin shell 
  formalism \cite{Anzhong}. A family of dynamic \wh\ solutions to the Einstein-Maxwell-scalar 
  equations was obtained in \cite{JinboYang}. General properties of arbitrary dynamic
  wormholes are discussed in \cite{Hayward2, Dynwh1}. 

  Overviews of various problems of wormhole physics can be found, for example, in 
  \cite{Visser, LoboRew}, see also the recent special issue of the Universe journal \cite{we-uni}.  
  
  In this paper we continue our study of possible traversable wormholes in GR sourced by
  such a classical and nonexotic source as dustlike matter, with or without an
  electromagnetic field \cite{IKS, we-wh21}. For electrically neutral dust, the general dynamic 
  spherically symmetric solution of GR was obtained by Lema\^{\i}tre and Tolman in 
  1933-1934 \cite{tolman,Lemaitre33} and later studied by Bondi 
  \cite{Bondi47,Christodoulou,Landau,Bambi}. It is generally called the Tolman or LTB solution. 
  The first attempt to construct a wormhole by selecting a special form 
  of arbitrary functions in this solution was made in \cite{IKS}. 

  An extension of the LTB solution including a radial \emag\ field was discussed in 
  \cite{markov,vickers,bailyn,shik,khlest,lapch} (see also references therein), where a complete 
  solution was achieved under some additional conditions, while in the general case, relevant  
  integrals of the Einstein-Maxwell equations were obtained and discussed. For arbitrary 
  electric charge distributions and arbitrary initial data, the problem was solved by Pavlov 
  \cite{pavlov}, and the solutions were further studied in \cite{pav-kb}; a further extension to 
  plane and hyperbolic symmetries of space-time were considered in \cite{kb83a, kb83b, kb84},
  see also references therein.
  
  In the present study we only consider configurations with an external magnetic (or electric) 
  fields and electrically neutral dust, however, if there is a \wh, its every entrance can comprise 
  a ``charge without charge'' \cite{Wheeler1, Wheeler2} due to electric or magnetic lines 
  of force threading the throat. Similar models with a special choice of initial data were studied 
  in \cite{Shatskii, hle-suh}, while here we do not restrict the initial data but consider the
  possible existence of such \whs\ in the cosmological context, being inscribed in Friedmann 
  models describing a matter-dominated stage of evolution, with possible inclusion
  of a cosmological constant which then describes dark energy. 

  Concerning dynamic wormholes, it is necessary to recall that there are different definitions 
  of dynamic \wh\ throats, which coincide with each other for static space-times, see, e.g., 
  \cite{Dynwh1, TomikawaIzumi, Bittencourt}. Following the papers \cite{Kar, Kim, Roman}, 
  we here choose the simplest definition based on the properties of 3-geometry of spatial 
  sections of space-time. This definition can in general be ambiguous due to the freedom 
  of choosing such spatial sections (or clock synchronization), but in the problem under
  consideration it looks most natural and intuitively clear. 
  
  The paper is organized as follows. In Section 2 we briefly describe the class of solutions 
  to be studied. In Section 3 we consider the conditions for possible existence of throats 
  and traversable wormholes. Section 4 describes a particular family of \wh\ solutions with 
  a wide enough range of parameters, to be used in Section 5 for placing them in the 
  cosmological context. The corresponding numerical estimates are obtained in Section 6, 
  indicating the possible existence and observable properties of such \whs\ in our universe.   
  Section 7 is a conclusion.

% ==========================================
\section{Extended LTB solution}
% ==========================================

  Let us consider a generalization of the original LTB solution \cite{tolman, Lemaitre33, Bondi47}, 
  describing the dynamics of a \sph\ distribution of electrically neutral dustlike matter in the 
  presence of an external electric or magnetic field and a cosmological constant. Let us, 
  for certainty, speak of a magnetic field which is more realistic astrophysically, keeping in mind
  that any further results can be easily re-interpreted in terms of an electric field. 
 
  If we choose a comoving reference frame for neutral dust particles, it is also a geodesic 
  reference frame for them, and the metric can be taken in the synchronous form
\beq                                                    \label{ds-tol}
      ds^2 = d\tau^2 - \e^{2\lambda(R,\tau)} dR^2 - r^2(R,\tau) d\Omega^2,
\eeq
  where $\tau$ is the proper time along the particle trajectories labeled
  by different values of the radial coordinate $R$, $\lambda(R,\tau)$ and $r(R,\tau)$ are 
  functions of $\tau$ and $R$. 
		
  The SET of dustlike matter reads $T_{\mu}^{\nu [d]}=\rho u_{\mu} u^{\nu}$, 
  where $\rho$ is the energy density and $u^{\nu}$ the velocity four-vector. The only nonzero
  component of this SET in the comoving reference frame, where $(u^{\nu})=(1, 0, 0, 0)$, is 
  $T^{0 [d]}_0 = \rho$. For the \emag\ field in the metric \rf{ds-tol}, the SET has the form
\beq   									                  \label{SET-e}
		{T\mN}^{[{\rm em}]} = \frac{q^2}{8\pi G r^4} \diag(1,\ 1,\ -1,\ -1),
\eeq  
  where $q$ can be interpreted as an electric or magnetic charge in suitable units   
  \cite{Reissner,Nordstrom}. 
  
  Nontrivial components of the Einstein equations with a cosmological constant  
  $\Lambda$ may be written as
\bearr 			\label{G11}
	   1+ 2r\ddot{r} + \dot{r}{}^2 - e^{-2\lambda}r'{}^2 = \frac{q^2}{r^2}
	   		+ \Lambda r^2,  
\\ \lal                 \label{G00}
	  1 + \dot{r}^2 + 2 r\dot{r}\dot{\lambda}
       	- \e^{-2\lambda} (2rr'' + r'{}^2 - 2 rr'\lambda' )
            							= 8\pi G \rho r^2 + \frac{q^2}{r^2} + \Lambda r^2,   
\yyy		\label{G01}
		      \dot{r}' - \dot{\lambda} r' = 0,
\ear
  where the dot stands for $\d/d t$ and the prime for $\d/d R$.
  The conservation law for dust matter, 
  $\nabla_\nu T_0^{\nu [d]}=0 \then \dot\rho + \rho(\dot\lambda + 2 \dot r/r) =0$, 
  leads to
\beq                 \label{rho}
		\rho  = \frac{1}{8\pi G} \frac{F'(R)}{r^2 r'}  \qq \Longleftrightarrow \qq
		F(R) = 8\pi G \int \rho r^2 r' dR,
\eeq  
  where $F(R)$ is an arbitrary function, which according to \rf{rho} may be said to describe 
  the initial mass distribution. On the other hand, \eq (\ref{G01}) is readily integrated 
  in $\tau$ with the result
\beq 
		\e^{2\lambda} = \frac{r'{}^2}{1 + f(R)},                  \label{lam}
\eeq
  where $f(R) > -1$ is one more arbitrary function. With (\ref{lam}), \eq (\ref{G11}) is
   rewritten as
\beq 							\label{ddr}
  			  2r \ddot{r} + \dot{r}^2 = f(R) + \frac{q^2}{r^2}+ \Lambda r^2,
\eeq
  and its first integral is
\beq 									 		\label{dr}
			\dot{r}^2 = f(R) + \frac{F(R)}{r}- \frac{q^2}{r^2} + \frac \Lambda 3 r^2,
\eeq
  where (as can be easily verified) the function $F(R)$ is the same as in \eq \rf{rho}.
  This expression reveals the physical meaning of $f(R)$ as a function characterizing the 
  initial radial velocity ($\dot r$) distribution of dust particles. Furthermore, if $\Lambda =0$,
  only under the condition $f \geq 0$ the particle can reach large values of $r$, so that 
  $f >0$ and $f=0$ correspond to hyperbolic and parabolic type of motion, respectively.
  In the case $f(R) <0$  (elliptic motion), the particle can at most reach a radius
  corresponding to the condition $\dot r=0$ in \eq \rf{dr}. With $\Lambda \ne 0$, things 
  are more involved, and the boundary of finite motion is shifted.
  
  Further integration of \eq \rf{dr} with $\Lambda \ne 0$ leads to elliptic integrals. In 
  what follows, for simplicity, we assume $\Lambda=0$, so that only elementary 
  functions are necessary to describe the solution (to be called for brevity the q-LTB solution). 
  Also, in what follows we will only need the description of elliptic motion, $f < 0$. In this case, 
  integration of \eq \rf{dr} gives 
\beq             \label{tau-}
     		 \pm[\tau-\tau_0(R)] =  \frac 1h \sqrt{-hr^2 +Fr -q^2} 
     		 			+ \frac {F}{2 h^{3/2}} \arcsin \frac {F-2hr}{\sqrt{F^2 - 4hq^2}}, 
\eeq
  where $h(R) := -f(R) > 0$, and $\tau_0(R)$ is one more arbitrary function that corresponds 
  to a choice of spatial sections of our space-time, or, in other words, to clock synchronization 
  between different dust layers with fixed values of $R$ (Lagrangian spheres).
  It is easy to see that elliptic motion is possible only with $F^2 - 4hq^2\geq 0$.
  
  For the solution \rf{tau-}, there is a convenient parametric representation
  (see, e.g., \cite{Landau, hle-suh}),
\bear                 			 \label{eta}
			r \eql \frac {F}{2h} (1 - \Delta \cos \eta), 
\nn
			\pm[\tau-\tau_0] \eql \frac {F}{2h^{3/2}}(\eta - \Delta \sin\eta),
						\qquad  \Delta = \sqrt{1 - \frac{4 hq^2}{F^2} },
\ear  
  where $0 < \Delta \leq 1$, and $\Delta =1$ corresponds to the original LTB solution 
  without an electromagnetic field. Notably, if $q \ne 0$, hence $\Delta < 1$, the model 
  has no singularities characterized by $r=0$, i.e, shrinking of a Lagrangian sphere to a
  point. Another kind of singularities, called shell-crossing or shell-sticking singularities and
  characterized by $r' =0$ while $F' \ne 0$ (see \rf{rho} and \rf{Kre}), are not excluded. 
  
  An important special case of the LTB solution ($q =0, \Delta =1$) is Friedmann's 
  closed isotropic cosmological model with dust matter, that corresponds to the following 
  choice of arbitrary functions \cite{Landau}: 
\beq       \label{Fri1}
	    F(\chi) = 2 a_0 \sin^3\chi, \qquad   h (\chi) = \sin^2 \chi, \qquad a_0 = \const
\eeq  
  (here, the radial coordinate $R = \chi$ is a ``radial angle'' on a 3D sphere), so that
\beq         \label{Fri2}  
  		r = r(\eta,\chi) = a(\eta) \sin \chi,  \qq  a(\eta) = a_0 (1-\cos \eta),
  		\qq  \tau = a_0 (\eta - \sin \eta),
\eeq  
  where $a(\eta)$ is the cosmological scale factor, and it is taken $\tau_0 =0$.
  
% =====================
\section{Possible throats}
% =====================
\def\th{_{\rm th}}

  As is clears from \rf{rho}, to keep the density positive, it is necessary to require 
  $F'/r' > 0$, but it does not mean that both $F'>0$ and $r'>0$. Therefore, one can
  admit the existence of regular maximum or minimum values of $r$ (at fixed $\tau$), 
  which can be interpreted as equators and throats, respectively. 
  
  As already mentioned, among different definitions of a \wh\ throat in dynamic space-times,
  we choose the definition \cite{Kar, Kim, Roman} according to which a {\bf throat} 
  in a space-time with the metric \rf{ds-tol} is a regular minimum of the spherical radius 
  $r(R, \tau)$ at a fixed value of $\tau$ (hence a fixed spatial section of space-time). Then, 
  as always, a wormhole is understood as a space-time region that contains a throat and 
  extends to sufficiently large $r(R, \tau)$ on both sides from this throat. 
  Further on we try to build \wh\ configurations based on the q-LTB solution. To do that, 
  let us first of all determine the conditions characterizing a \wh\ throat \cite{we-wh21}.
  
  The 3D spatial metric of a spatial section $\tau = \const$ is
\beq                              \label{ds_3}
           dl_{(3)}^2 = \frac{r'{}^2 dR^2}{1+ f(R)} + r^2(R) d\Omega^2.
\eeq   
  where $r(R)=r(R,\tau)\big|_{\tau=\const}$, and the coordinate $R$ is still arbitrary. 
  To formulate  the throat conditions, let us choose the manifestly admissible Gaussian 
  coordinate $l$, measuring length in the radial direction, such that 
  $dl = |g_{RR}|^{1/2} dR$. Then at a throat we must have 
\beq                              \label{th}
			\frac{dr}{dl} = 0, \qquad  \frac{d^2 r}{dl^2} > 0 
\eeq  
  (for a generic minimum of $r$, ignoring possible high-order ones, with $d^2 r/dl^2 =0$).  
  From the first condition it follows that on the throat, $R= R\th $,
\beq                              \label{th1}
			\frac{dr}{dl} = \sqrt{1+ f(R\th )} = 0 \ \ \then \ \ f(R\th ) = -1, 
					\ \ \ {\rm or}\ \ \  h(R\th )=1.
\eeq      
  Thus it is clear that only elliptic models \rf{tau-} are compatible with \wh\ existence.
  Next, to keep the metric \rf{ds-tol} nondegenerate, it must be in general $1+f  = 1 -h >0$, 
  while $h = 1$ is only admissible at a particular value of $R$, therefore, $R=R\th $ should 
  be a maximum of $h(R)$, such that $h'(R\th ) =0$ and $h''(R\th ) <0$. Then the second 
  condition \rf{th} implies
\beq                             \label{th2}
			\frac{d^2 r}{dl^2}\Big|_{R=R\th } = -\frac{h'}{2r'}\Big|_{R=R\th } > 0.
\eeq   
  Thus $h'(R)$ vanishes at $R=R\th $ together with $r'(R)$, with a finite limit of 
  their ratio. The conditions (\ref{th1}) and (\ref{th2}) lead to restrictions on the 
  arbitrary functions $F(R)$ and $h(R)$. 
  
  As follows from \rf{rho}, the dust density tends to infinity, thus indicating a singularity, if 
  either $r\to 0$ or $r' \to 0$, except for cases where both $r'\to 0$ and $F' \to 0$ at finite $r$, 
  keeping finite the ratio $F'/r'$, precisely what happens at a \wh\ throat. That the space-time 
  remains regular under these circumstances, can be confirmed by calculating
  the Kretschman scalar $\cK$,
\beq                       \label{Kre}
		\cK(R,t) =3\frac{{F'}^2}{{r'}^2r^4}-8\frac{F'F}{r'r^5}+12\frac{F^2}{r^6}
					+20\frac{F'q^2}{r'r^6} - 48\frac{Fq^2}{r^7}+56\frac{q^4}{r^8}.
\eeq
  Thus, at possible throats, all three derivatives $r'$, $F'$ and $h'$ vanish, with finite limits of 
  their ratios.
  
  From \rf{eta}, we obtain the following expression for the derivative $r'$ on a 
  constant-$\tau$ section of our space-time:
\bearr 
		r'=\frac{Fh'N_1(R,\eta) + 2h F' N_2(R,\eta)} {4\Delta h^2(1-\Delta\cos\eta)}, 
\nnnv
		N_1(R,\eta) = \cos\eta -3\Delta + 3\Delta^2 (\eta\sin\eta+\cos\eta) 
							 + \Delta^3 ( -2 + \cos^2\eta),
\nnnv					
		N_2(R,\eta) = 	- \cos\eta + 2\Delta- \Delta^2(\cos\eta + \eta\sin\eta).\label{r_R}
\ear
  At a throat $R=R\th $, the ratios $F'/r'$ and $h'/r'$ are finite and nonzero 
  (though with different signs), $r'$, $h'$ and $F'$ are there small quantities of the same order 
  of magnitude.
  
  We can summarize the throat conditions as follows:
\bearr                    \label{throat}
		h = 1, \quad\ h'=0, \quad\ h'' < 0,
\nnn		
		F' = 0, \quad\ r' =0,\quad\   \frac{h'}{r'} < 0, \quad\  \frac{F'}{r'} > 0. 
\ear
   Also, we have everywhere $F^2 - 4hq^2 > 0$ and $\Delta \leq 1$.

   For further analysis, let us consider the limit $\lim\limits_{R\to R\th }\dfrac{Fh'}{F'h}=-B$ 
  such that $B = \const  \geq 0$. Then for $r'$ near the throat we obtain 
\bearr 
		r'\Big|_{R\to R\th } \approx \frac{F'(2N_2 - B N_1)} {4\Delta (1-\Delta\cos\eta)},
\ear
  It vanishes either if $F'=0$ or if $N_* = 2N_2 - BN_1 = 0$. The density (\ref{rho}) 
  on the throat is given by
\bearr          \label{rho_th}
		\rho(R\th ,\eta) =  \frac {\Delta (1 - \Delta\cos\eta)} 
		{2\pi G r^2 (2N_2-BN_1)}\bigg|_{R\th },
%\\ \lal                    \label{d2r_th}
%		\frac{d^2 r}{dl^2}\bigg|_{R\th } =  \frac{2 B}{F} 
%					\frac{\Delta( 1-\Delta\,\cos \eta)}{2N_2-BN_1}\bigg|_{R\th }.
\ear
  and it blows up where $N_* = 2N_2 - BN_1 = 0$ while the other factors are positive 
  ($F > 0$ by assumption). Meanwhile, $N_*$ has different signs at the ends and the 
  middle of the range of $\eta$: %(see also Fig.\,\ref{N*}):
\bearr
	N_*\Big|_{\eta=0,2\pi} = - (1-\Delta)^2 [2+B (1-\Delta)] \leq 0, 
\nnn	
	N_*\Big|_{\eta=\pi} = (1+\Delta)^2 [2+B (1+\Delta)] > 0.
\ear
  Therefore, we inevitably obtain $N_* = 0$, hence a singularity, at (at least) two values of 
  $\eta$ say, $\eta_1< \pi$ and $\eta_2> \pi$, for any $\Delta < 1$ ($q \ne 0$). These are 
  so-called shell-crossing singularities forming due to $r'\to 0$ while $r$ is finite.
  
  In the case $q=0$, $\Delta=1$ (pure dust), we see that $N_*$ vanishes at 
  $\eta_{1,2} =0,\,2\pi$ and is positive at $\eta \in (0, 2\pi)$. 

  Thus a nonsingular evolution period for a throat $R=R\th $, with finite density $\rho > 0$, 
  takes place at times $\eta_1 < \eta < \eta_2$ at which $N_* > 0$. For other Lagrangian 
  spheres $R=\const$ we obtain similar but other time limits due to $R$ dependence of 
  the functions $F$ and $h$. 
  
  The above relations lead to general restrictions on the dust densities in the \wh\ solutions.
  For example, consider the solution with $q=0$ at $\eta = \pi$, that is, at maximum expansion.
  In this case, $r = F(R)/h(R)$, and $2 N_2 = - N_1 =8$, and according to \rf{rho} we obtain
\beq  			\label{rho-pi}
		\rho\Big|_{\eta=\pi} = \frac{1}{8\pi G}\frac{F'}{r^2 r'} 
		   = \frac{1}{8\pi G}\frac{F' h^4}{F^2 (F' h - Fh')}
		   = \frac{h}{8\pi G r^2 (\big(1 - rh'/F'\big)}.
\eeq		   
  In all \wh\ solutions, $h \leq 1$; furthermore, $h'/F' < 0$ near the throat, and let us suppose 
  that this is also true at other values of $R$ ($F'<0$ at $r'>0$ would give negative matter 
  densities; while a changing sign of $h'$ is still possible). Then \rf{rho-pi} leads to the 
  simple inequality
\beq             \label{rho-ineq}
	\rho \leq \frac{1}{8\pi G r^2} \approx 6.8\ten{26} {\rm \frac {g}{cm^3}}\frac{\rm cm^2}{r^2}.
\eeq
  For example, at the throat we have $h =1$ and $-rh'/F' = B >0$. This inequality actually admits
  very large density values: thus, if the throat radius is 1 km, we have the restriction 
  $\rho \lesssim 10^{16}\,\rm g/cm^3$, a supernuclear density, hard to imagine with dustlike matter. 
  We can also notice that the throat density values are diminished by large values of $B$. 
    
% --------------------------------- fig 1, r_throat 
\begin{figure}
\centering
\includegraphics[scale=0.35]{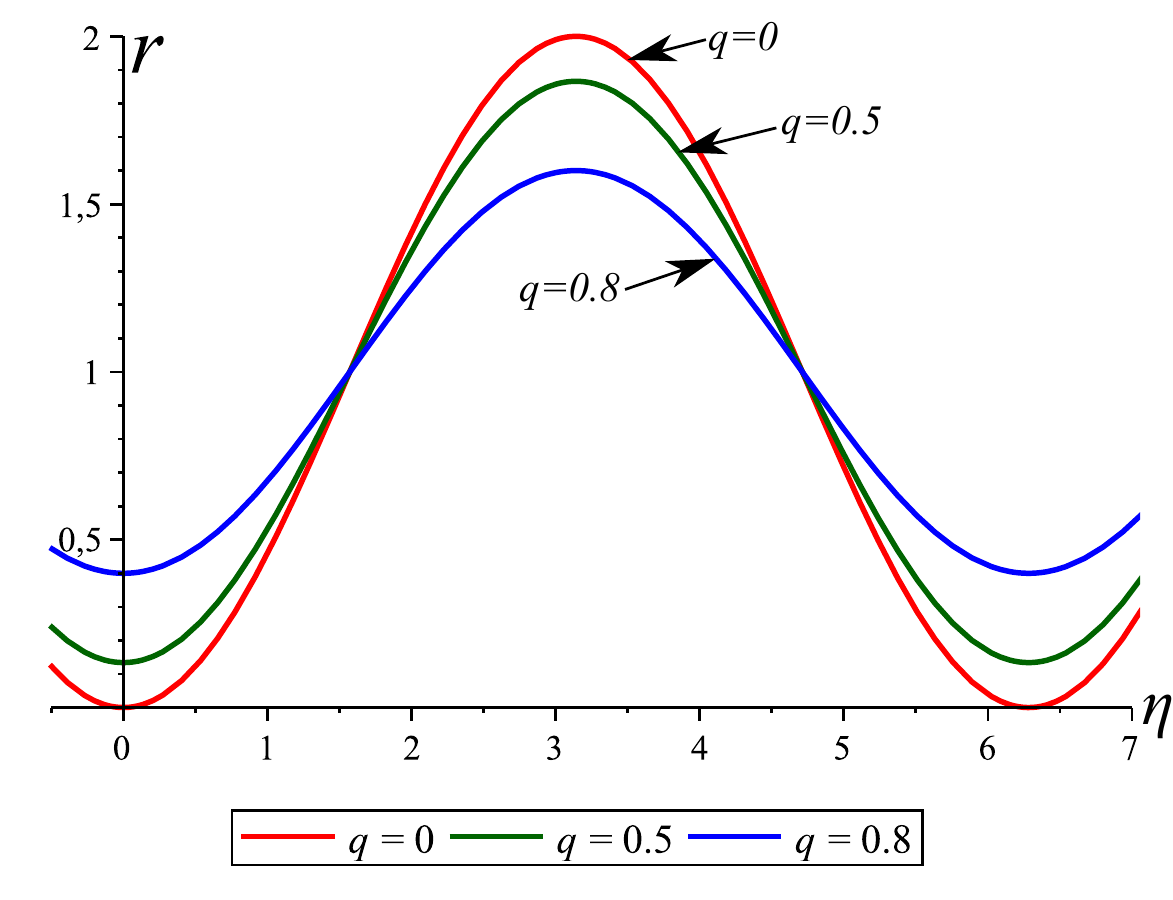}\ 
\includegraphics[scale=0.35]{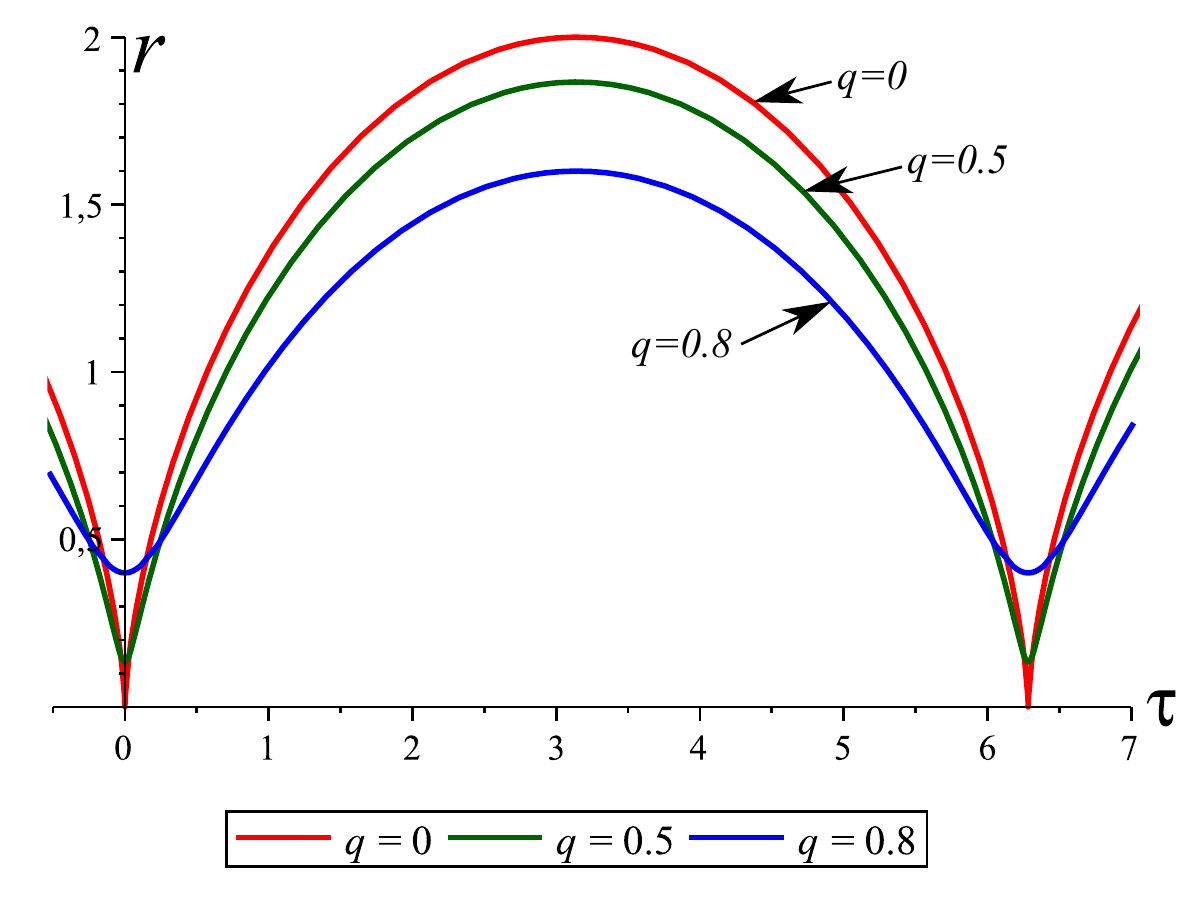}
\caption{\small
	Time dependence of the throat radius for $q=0, 0.5, 0.8$ in terms of $\eta$ (left) 
	and in terms of $\tau$ (right). \label{fig2}}
\end{figure}
% ----------------------------------  fig 2, rho_throat
\begin{figure}
\centering
\includegraphics[scale=0.5]{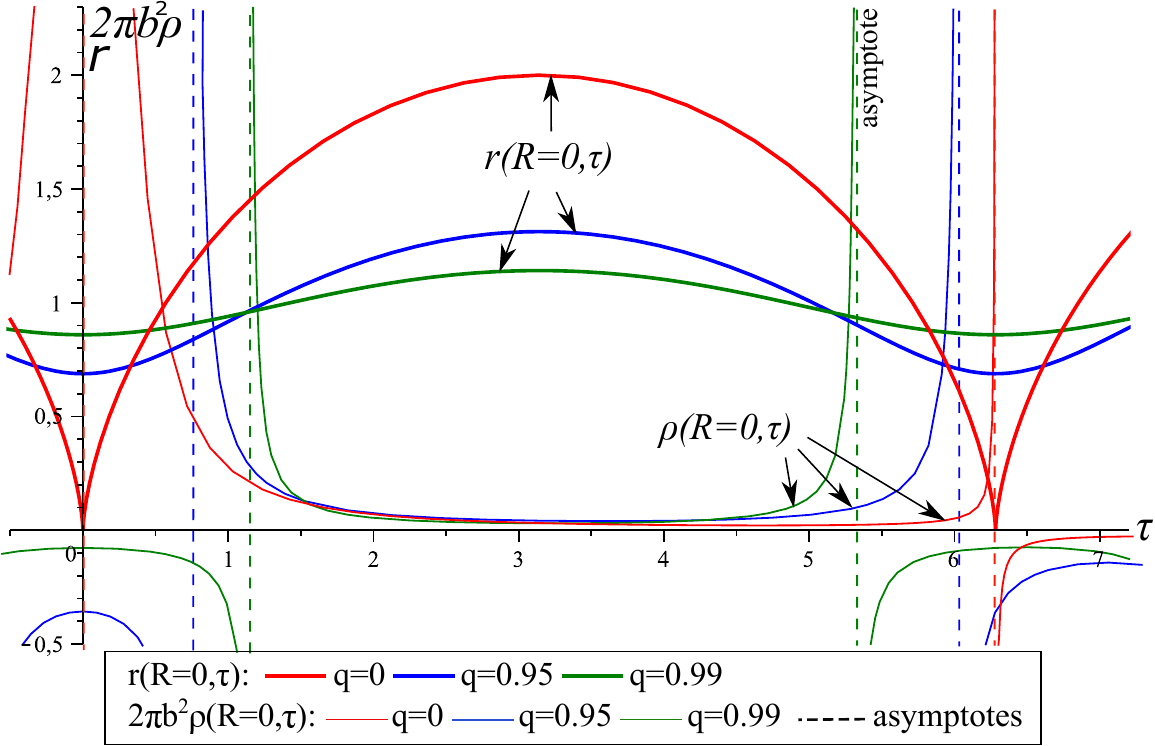}
\caption{\small
	Time dependence of the functions $2\pi b^2\rho$ (thin lines) and $r(0,\tau)$ (thick lines) 
	on the throat $R=0$ for $q=0,\,0.95,\,0.99$ in the model \rf{ex1}. 
	Dashed lines show the asymptotes. For other values of $q$ the plots look in a similar way. 
	\label{rho,r_throat}}
\end{figure}   
% --------------------------------------

% =====================================
\section{A particular family of \wh\ solutions}
% =====================================

  Let us select a family of LTB \wh\ solutions, choosing the following 
  simple functions of $R$ in agreement with the requirements \rf{throat}:
\beq                												 \label{ex1}
		h(R)=\frac{1}{1+R^2}, \quad \ F(R)=2b (1+R^2)^k ,
		\ \ \then \ \ 
		\Delta=\sqrt{1-\frac{q^2}{b^2(1+R^2)^{2k+1}}}, 
\eeq		
  with the constants $b, k > 0$. This choice of $h(R)$ is made without loss of generality due to 
  arbitrariness of the $R$ coordinate, while the choice of $F(R)$ is significant.
  In particular, since both $h(R)$ and $F(R)$ are even functions, the \wh\ is symmetric 
  with respect to its throat $R=0$ In \rf{ex1}, the constant $b$ specifies a length scale, and we have
\bearr                												\label{ex1-r}
		r(R,\eta) = b(1+R^2)^{k+1}\, (1 - \Delta \cos \eta),\qq    
		\tau-\tau_0(R) = b(1+R^2)^{k+3/2}\, (\eta - \Delta \sin \eta) 
\yyy
		r'(R,\eta) = \frac{bR (1+R^2)^k (2k N_2 - N_1)}{\Delta(1 - \Delta \cos \eta)}, 
\ear
  with $N_{1,2}$ defined in \eqn{r_R}. The density $\rho$ (\ref{rho}) and the quantity 
  $d^2r/dl^2$ at $R=0$ then read
\bear  			\label{rho_th_ex}
	\rho(R,\eta) \eql \frac{k\Delta}{2\pi G b^2 (1+R^2)^{2k+3} (1-\Delta\cos\eta) (2kN_2-N_1)},
\\                        \label{d2r_th_ex}
	\frac{d^2r}{dl^2}\bigg|_{R=0} \eql \frac{\Delta\,( 1-\Delta\,\cos \eta)}{b(2kN_2-N_1)}\bigg|_{R=0}.
\ear
  As already noted, different signs of the derivatives of $h(R)$ and $f(R)$, under the condition
  $2N_2(R,\eta) - N_1(R,\eta) > 0$, provide the validity of the throat conditions \rf{throat} at $R=0$
  and, by continuity, in some its neighborhood, but the same is not guaranteed at all $R$ and $\eta$. 
  
  {The time dependence of the throat radius $r\th$ and the density $\rho\th$ on the throat 
  was studied in \cite{we-wh21}. Here, for completeness, we reproduce some figures 
  from \cite{we-wh21}. Thus, Fig.\,\ref{fig2} shows the time dependence of the throat radius, 
  while the density $\rho\th$ is shown in Fig.\,\ref{rho,r_throat} for $k =1$ and different 
  values of $q$, where dashed lines show the asymptotes of the function.}
  Finite positive density values are observed for a limited period of time $\eta \in (\eta_1, \eta_2)$ 
  while $2N_2-N_1 >0$, between two singularities where $\rho$ and $\cK$ diverge. Outside this 
  interval, in the case $q\neq 0$, the density changes its sign along with $d^2r/dl^2$, therefore 
  the throat conditions hold together with the condition $\rho > 0$. 

  Thus we observe a good \wh\ behavior of our solution at the time interval $\eta \in (\eta_1, \eta_2)$. 
  With decreasing charge, this interval increases; and at $q=0$ we have $\eta_1=0$, $\eta_2=2\pi$. 
  Outside the throat (at $R\neq 0$), the plots look similarly, but the singularities occur at other 
  time instants.

% ==========================================
\section{Matching to a dust-filled Friedmann universe}
% ==========================================
\subsection{General observations}
% ---------------------------------------------

   Now, let us look how the \wh\ solution discussed above can be inscribed into the closed 
   Friedmann isotropic space-time characterized by the relations \rf{Fri1}, \rf{Fri2}. At that, 
   we can note \cite{we-wh21} that to join two LTB space-time regions, characterized by 
   different functions $F(R)$ and $h(R)$, at some hypersurface $\Sigma$ corresponding to a fixed 
   value of the radial coordinate $R = R^*$, one should first of all identify  $\Sigma$ as viewed 
   from different sides. Hence the metric tensor must be continuous on $\Sigma$. With the 
   metric \rf{ds-tol} it simply leads to $[r^2(R,\tau)] =0$ (as usual, square brackets denote jumps 
   when crossing the transition surface $\Sigma$), while by \rf{ds-tol} $g_{\tau\tau} \equiv 1$ on 
   both sides and does not lead to any further requirements.
   
   Next, to avoid the emergence of a shell of matter on the junction surface $\Sigma$, according to 
   the Darmois-Israel matching conditions \cite{darmois, israel}, one should require continuity of
   the second quadratic form on $\Sigma$. When applied to the metric \rf{ds-tol}, this requirement 
   leads to $[\e^{-\lambda}g'_{\tau\tau}] =0$ (which holds trivially due to $g_{\tau\tau}\equiv 1$) 
   and $[\e^{-\lambda} r'] =0$.  As a result, with \rf{lam} and \rf{eta}, we obtain
\beq                          \label{junc}
           [r] = 0, \quad\  [\e^{-\lambda} r'] =0 \ \ \then \ \ [h] =0, \quad [F]=0..
\eeq    
  Thus to match two LTB solutions on a surface $\Sigma$ ($R=R^*$), it is sufficient to
  identify the values of $F(R^*)$ and $h(R^*)$ in these solutions. It is important that by 
  \eq \rf{eta} the above matching conditions hold at all times at which both solutions remain 
  regular. Also, there is no necessity to worry about the choice of the radial
  coordinates on different sides of $\Sigma$ because both quantities $r$ and $\e^{-\lambda} r'$ 
  are insensitive to the choice of the coordinate $R$, and at reparametrizations of $R$ the 
  arbitrary functions $h(R)$ and $F(R)$ behave as scalars and preserve their values.
  
  Now, let us apply the conditions \rf{junc} to the Friedmann solution \rf{Fri1}, \rf{Fri2} with $q=0$ 
  and an arbitrary \wh\ solution  described above, also putting $q=0$, and let us specify the 
  junction surface $\Sigma$ by some values of the radial coordinates $\chi = \chi_*$ and 
  $R = R_* >0$ (here and henceforth we mark by an asterisk the values of different quantities 
  on $\Sigma$). We then obtain for the \wh\ solution
\beq                    \label{junc1}
		 h_* = \sin^2 \chi_*, \qq F_* = 2a_0 \sin^3 \chi_* \ \then \ F_* = 2a_0 h_*^{3/2}.			
\eeq  
  Consider, as before, the instant of maximum expansion, $\eta =\pi$, then $r = F/h$, and 
  according to \rf{junc1} we obtain
\beq                    \label{junc2}
			F_* = r_*h_* = 2a_0 h_*^{3/2}  \ \then \ h_* = r_*^2/(4 a_0^2).
\eeq  
  For the density we can apply \eqn{rho-pi}, hence on $\Sigma$ we have 
\beq                    \label{rho-j}
         		\rho_*\Big|_{\eta=\pi} = \frac{h_*}{8\pi G r_*^2 \big(1 - r_*h'_*/F'_*\big)}
         				= \frac 1 {32\pi G a_0^2 \big(1 - r_*h'_*/F'_*\big)}.           				
\eeq  
  Assuming, as before, that everywhere in the \wh\ solution $F'/h' < 0$, we arrive at the inequality
\beq                    \label{rho-j-ineq}
         		\rho_*\Big|_{\eta=\pi} <  \frac 1 {32\pi G a_0^2} \approx 5.5\ten{-30} \rm\frac{g}{cm^3}
\eeq  
  if we assume $a_0 \sim 10^{28}$ cm, approximately the size of the visible part of the Universe.
  
  On the other hand, in the Friedmann solution \rf{Fri1}, \rf{Fri2} the matter density is 
\beq                    \label{pho-Fr}
		\rho_{\rm Fr}(\eta) = \frac {3}{4\pi G a_0^2 (1-\cos\eta)^3}, \qq
				\rho_{\rm Fr}\Big|_{\eta=\pi} = \frac {3}{32\pi G a_0^2}.
\eeq  
  Thus according to \rf{rho-j-ineq}, the \wh\ matter density at the junction surface $\Sigma$
  is not only very small, but it is even a few times smaller (by at least a factor of three) than the 
  cosmological matter density.  In other words, the \wh\ region is, at least close to $\Sigma$, 
  a region of smaller density, maybe resembling a void. This observation was made for the 
  instant $\eta =\pi$, but it remains true at all times since the $\eta$ dependence is the 
  same for the \wh\ and cosmological solutions. 

  Some more general observations can be made. As follows from the throat conditions \rf{throat}, 
  $h(R)$ has there a maximum with $h=h_{\rm th} = 1$, while $F(R)$ has a minimum, therefore, 
  according to \rf{eta},
\beq
		F(R)\geq  F(0) = r_{\rm th}\Big|_{\eta=\pi},				\label{_eq1}
\eeq
  where, for simplicity, $h(R)$ and $F(R)$ are assumed to be monotonic in the ranges $R>0$ and $R<0$.
  Considering, as before, the instant of maximum expansion, $\eta =\pi$, from \eqs (\ref{junc1}), 
  (\ref{junc2}) we obtain at the junction surface $R=R_*$:
\beq
		F_*=2a_0h_*^{3/2},\quad r_*\Big|_{\eta=\pi}=\frac{F_*}{h_*}
 \ \then \ 
		r_*=F_*^{1/3}(2a_0)^{2/3},\label{_eq2}
\eeq
  Then from (\ref{_eq1}) and (\ref{_eq2}) it follows
\bearr                     \label{_eq3}
	r_*\geqslant (2a_0)^{2/3}r_{\rm th}^{1/3}=7.4\times 10^{18} \, 
				\left(\frac{r_{\rm th}}{{\rm cm}}\right)^{1/3} \,  {\rm cm}, 
\yyy				
		 r_{\rm th}\leqslant   \frac{r_*^3}{(2a_0)^2}
		 =2.5\times 10^{-57} \, \frac{r_*^3}{{\rm cm^3}} \, {\rm cm},
\ear
  where $r_{\rm th}$ and $r_{*}$ are taken at maximum expansion, $\eta =\pi$. We see that 
  the length scales of the wormhole region $r_*$ and its throat $r_{\rm th}$ are substantially different 
  in cases of physically interest, $r_* \ll a_0$. Furthermore, the throat lifetime is 
  $\Delta\tau_{\rm th}=2\pi r_{\rm th}/c$, while the lifetime of the wormhole region coincides 
  with that of the universe, $\Delta\tau_*= 2\pi a_0 /c \approx 2\times 10^{18}$ s, and thus we 
  have $\Delta\tau_{\rm th}\ll \Delta\tau_*$. All these estimates (\ref{rho-ineq}), (\ref{rho-j-ineq}) 
  and (\ref{_eq3}) are based on our general assumptions about the model. Numerical estimates 
  for a specific choice of the functions $h(R)$ and $F(R)$ will be made below.

% -----------------------------------------------------------
\subsection{Estimates for a particular model}
% -----------------------------------------------------------

  Now, to obtain further estimates, let us describe the wormhole region by Eqs. \rf{ex1}, \rf{ex1-r},
  then the junction conditions \rf{junc} lead to  
\begin{equation}             \label{junc-k}
		R_* = \cot \chi_*, \quad   b = a_0 (\sin\chi_*)^{3+2k},
\end{equation}   
  which provides matching at $R^* > 0$. Since the functions involved in \rf{ex1} are even, 
  a similar kind of matching can be applied at $R^* < 0$. The whole composite model
  then consists of two closed evolving dust-filled Friedmann universes, connected 
  through a \wh, thus forming a dumbbell-like configuration, or otherwise we can suppose
  that negative values of $R$ lead to the same Friedmann universe at some different location.
  
  Some numerical estimates are in order. Taking, as before, $a_0 \sim 10^{28}$ cm, 
  let us also assume that the \wh\ region is small as compared to the whole universe, 
  hence, $\chi_* \ll 1$, and $\sin \chi_* \approx \chi_*$. Accordingly,
\begin{equation}   
		 R_* = 1/\chi_*, \quad
		 h_* = \chi_*^2, \quad 
		 F_* = 2b \chi_*^{-2k}, \quad   
		 b = a_0 \chi_*^{2k+3}.
\end{equation}   
  Note that in the \wh\ solution $r(R,\eta) = b(1+R^2)(1-\cos\eta)$, and $R=0$ is the throat,
  so $2b$ is the maximum value of the throat radius, $2b = r(0, \pi)$.
  
  Relationships for the wormhole parameters are easily calculated. Equations \rf{rho-pi} and \rf{ex1} 
  imply $r h'/F'=-1/k$, and we get for the matter density
\bearr  
    \rho_{\rm th}=\frac{c^2}{32\pi G b^2}\frac{k}{k+1}
       \approx 1.3\times 10^{26}  \frac{k}{k+1} \frac{\rm g}{\rm cm^3} \frac{{\rm cm^2}}{b^2}, 
\yyy
	 \rho_*=\frac{c^2}{32\pi G a_0^2}\frac{k}{k+1}
		\approx 1.3\times 10^{-30} \frac{k}{k+1}\frac{\rm g}{\rm cm^3}.
\ear 
  The junction conditions (\ref{junc-k}) imply
\beq
		r_*=2a_0\left(\frac{b}{a_0}\right)^{1/(2k+3)}.
\eeq
  The minimum value of $r_*$ for given $b$ corresponds to the limit $k\to 0$, specifically. 
  $r_*\geqslant 2a_0^{2/3}b^{1/3}$ (\ref{_eq3}).
  
  Tables \ref{table1} and \ref{table2} show some estimates of the \wh\ parameters, such as the 
  throat radius $r_{\rm th}=2b$, matter density $\rho_{\rm th}$ on the throat and the radius $r_*$ 
  of the whole wormhole region in the cases $k=0.1$ and $k=1$. The density at the junction surface 
  does not depend on $b$ and equals $\rho_*=1.2\times 10^{-31}\,{\rm g}/{\rm cm}^3$ for $k=0.1$, 
  and $\rho_*=6.7\times 10^{-31}\,{\rm g}/{\rm cm}^3$ for $k=1$. We see that the wormhole 
  region has the size of parsecs or more even for small throats. Near the throat, the density is 
  super-nuclear for $b = 1$\,km, it is of white-dwarf order near a throat of planetary size, and 
  reasonably small near a throat of 1 pc. At the junction, the density $\rho_*$ is smaller than the 
  mean cosmological density, as should be the case according to our general observations.

% -------------------------------- tab 1
\begin{table}
\centering
\caption{\small
	Estimates of matter density $\rho_{\rm th}$ at the throat and the radius $r_*$ of the wormhole 
	region for different throat radii $r_{\rm th}$, in the case $k=0.1$,
	$\rho_*=1.2 \times 10^{-31}\,{\rm g}/{\rm cm}^3$. }
\medskip
\begin{tabular}{|c|c|c|c|}
\hline
 $r_{\rm th}$  &  $r_*$   & $\rho_{\rm th}$ ${\rm [g/cm^3]}$   \\ 
\hline\hline
 $1.6\times 10^{-33}$ cm (Planck length) & $1.6\times 10^{4}$ km (Earth) &$1.9\times 10^{91}$   \\
\hline
  $1$ km  & $10^{21}\,{\rm cm}=338$ pc  & $4.9\times 10^{15}$ (nuclear density) \\
\hline
 $10$ km (neutron star)  & $700$ pc &$4.8\times 10^{13}$  \\
\hline
$6.4\times 10^3$ km (Earth)  & $5.1$ Kpc  & $1.4\times 10^{8}$ (white dwarf) \\
\hline
 $2.3\times 10^{5}$ km &$16$ Kpc (Milky Way) & $94\times 10^{3}$  \\
\hline
 $695\times 10^{3}$ km (Sun) & $23$ Kpc & $10^4$  \\
\hline
  $10^{7}$ km (super BH) & $52$ Kpc & $49$ \\
\hline
  $7\times 10^{7}$ km & $96$ Kpc & $1$ (${\rm H_2O}$)  \\
\hline
  $1$ pc & $5.7$ Mpc & $5.1\times 10^{-12}$  \\
\hline
 $6.5$ pc & $10$ Mpc (galaxy cluster) & $1.9\times 10^{-13}$  \\
\hline
 $10$ Kpc &$100$ Mpc (void) & $4.8\times 10^{-20}$ (interstellar medium) \\
\hline
\end{tabular}
\label{table1}
\end{table}
% --------------------------------------------------- tab 2
\begin{table}
\centering
\caption{\small
	Estimates of matter density $\rho_{\rm th}$ at the throat and the radius $r_*$ of the wormhole 
	region for different throat radii $r_{\rm th}$, in the case $k=1$, 
	$\rho_*=6.7\times 10^{-31}\,{\rm g}/{\rm cm}^3$. }
\medskip
\begin{tabular}{|c|c|c|c|}
\hline
 $r_{\rm th}$  &  $r_*$   & $\rho_{\rm th}$ ${\rm [g/cm^3]}$   \\ 
\hline\hline
 $1.6\times 10^{-33}$ cm (Planck length) & $1.2\times 10^{11}$ km &$10^{92}$   \\
\hline
$2$ cm & $16$ Kpc (Milky Way) & $7\times 10^{25}$  \\
\hline
  $1$ km  & $0.14$ Mpc  & $2.7\times 10^{16}$ (nuclear density) \\
\hline
 $10$ km (neutron star)  & $0.2$ Mpc &$2.7\times 10^{14}$  \\
\hline
$1.6\times 10^4$ km  & $1$ Mpc  & $10^{8}$ (white dwarf) \\
\hline
  $695\times 10^{3}$ km (Sun) & $2.1$ Mpc & $5.6\times 10^4$  \\
\hline
  $10^{7}$ km (super BH) & $3.6$ Mpc & $268$ \\
\hline
  $1.6\times 10^{8}$ km & $6.2$ Mpc & $1$ (${\rm H_2O}$)  \\
\hline
 $1.7\times 10^{9} \, {\rm km}$  & $10$ Mpc (galaxy cluster) & $8.6\times 10^{-3}$  \\
\hline
  $1$ pc & $71$ Mpc & $2.8\times 10^{-11}$  \\
\hline
 $6.7$ pc & $100$ Mpc (void) & $8.6\times 10^{-13}$ (interstellar medium) \\
\hline
\end{tabular}
\label{table2}
\end{table} 
% ------------------------------------------------	
	
% ====================================
\section{Wormhole lifetime and traversability}
% ====================================

  Now we would like to consider the radial motion of photons in the model (\ref{ex1}) of a dust 
  layer, assuming that it is bounded by $|R| < R_*$ and is located between two copies of 
  Friedmann space-time. It is clear that a photon radially falling to such a \wh\ and reaching 
  the throat has no other way than to travel further in the direction of another universe or maybe 
  a distant part of the same universe. The question is whether or not it will go out from the 
  dust layer in this ``other'' universe rather than a singularity. In other words, is the \wh\ 
  (or the \wh\ part of space-time) traversable.
	
  Further on we will consider the motion of photons under different choices of the arbitrary function  	
  $\tau_0(R)$ in the solution \rf{eta} with $\Delta =1$ while in the Friedmann solution we fix
  $\tau_0 \equiv 0$. It should be noted here that for a particular LTB solution taken separately, 
  the choice of $\tau_0(R)$ means nothing else than clock synchronization, or, in other words, 
  the choice of spatial sections of the same space-time in the same reference frame. However, 
  in a composite model like ours, unifying two different LTB solutions, this choice is more meaningful, 
  and different $\tau_0(R)$ corresponds to different synchronization of events in one region relative 
  to events in the other region. Thus, fixing $\tau_0(R) \equiv 0$ in the \wh\ solution, we make 
  a physical assumption that the \wh\ throat emerges simultaneously with the whole Friedmann 
  universe, while $\tau_0(R)> 0$ means that  this happens later from the viewpoint of an observer 
  located in this universe. We will consider both options.

% -----------------------------------------------------------------------------------
\subsection{Radial motion of photons in the case $\tau_0= 0$}
% -----------------------------------------------------------------------------------

  From the metric (\ref{ds-tol}) it follows for null radial geodesics that
\bear
 		\frac{dR}{d\tau}=\pm\frac{\sqrt{1-h}}{|r'|},				\label{eq_ph}
\ear
  where the derivative $r'=r'(R,\tau)$ can be found from \eq (\ref{ex1-r}):
\bear
	r'(R,\tau)=
	\frac{b R (R^{2}+1)^{k}  }{1-\cos\eta } 
	\left[4k +5 -4(k +1) \cos \eta-(2k +3) \eta \sin \eta -\cos^{2}\eta\right].   \label{eq01}
\ear
  The plus sign in \eq (\ref{eq_ph}) corresponds to the photon motion through the \wh\ from 
  $R<0$ to $R>0$, and the minus sign to the opposite motion. Due to the symmetry of the 
  model, it is sufficient to consider, for example, the plus sign. 

  Let us calculate the time derivative of the spherical radius $r(\tau,R(\tau))$ along a light ray 
  $R=R(\tau)$:
\bear
	 \frac{d}{d\tau}r(\tau,R(\tau))=\frac{\d r}{\d\tau}+\frac{\d r}{\d R}\frac{dR}{d\tau}
	 =  \pm\sqrt{f+\frac{F}{r}}\pm \sqrt{1-h},\label{dr_dtau},
\ear
  where $R=R(\tau)$ describes the radial motion of a photon (\ref{eq_ph}), the first $\pm$ sign
  corresponds to the expansion (+) of the dust shells at $\eta\in(0;\pi)$ or their contraction ($-$)
  at $\eta\in(\pi;2\pi)$; {the second $\pm$ sign corresponds to photon motion from the 
  throat ($+$), or to the throat ($-$).} 	
  It is clear that $dr/d\tau<0$ or $>0$ means convergence or divergence of light 
  rays. Note that at the throat, $h(0)=1$, the photons move parallel to dust particles: 
  $dr(\tau,R(\tau))/d\tau = \d r/\d\tau$. 

  It is instructive to define an apparent horizon as the location of turning points for radial light rays, 
  that is, the set of events where the light rays stop diverging and start to converge, or vice versa,
  hence, 
\beq
		\frac{d}{d\tau}r(\tau,R(\tau))=0.					\label{app_hor}
\eeq
{  Using \eqs (\ref{ex1-r}) and (\ref{dr_dtau}), \eq (\ref{app_hor}) is rewritten in the form}
\beq
		\cot\frac{\eta}{2}\pm R=0,\label{app_hor2}
\eeq
  and finally we have the parametric equations for the apparent horizon
\bear
	{	r=\frac{2^{k+1}b}{(1-\cos\eta)^k},\qquad 
		\tau=\frac{2^{k+3/2}b(\eta-\sin\eta)}{(1-\cos\eta)^{k+3/2}}.}
\ear
{The condition (\ref{app_hor}) can be satisfied only if the terms in \eq (\ref{app_hor2}) have 
   different signs. Thus at the expansion stage $\eta\in(0;\pi)$ there is an apparent horizon for 
   photons moving towards the throat ($\pm R<0$), while at the contraction stage $\eta\in(\pi;2\pi)$, 
   on the contrary, for photons moving from the throat ($\pm R>0$). There are actually two 
   apparent horizons, depending on the direction of motion.}

% --------------------------------------------- fig 3
\begin{figure}[h]
    \begin{minipage}{0.5\textwidth}\centering
    \includegraphics[scale=0.4]{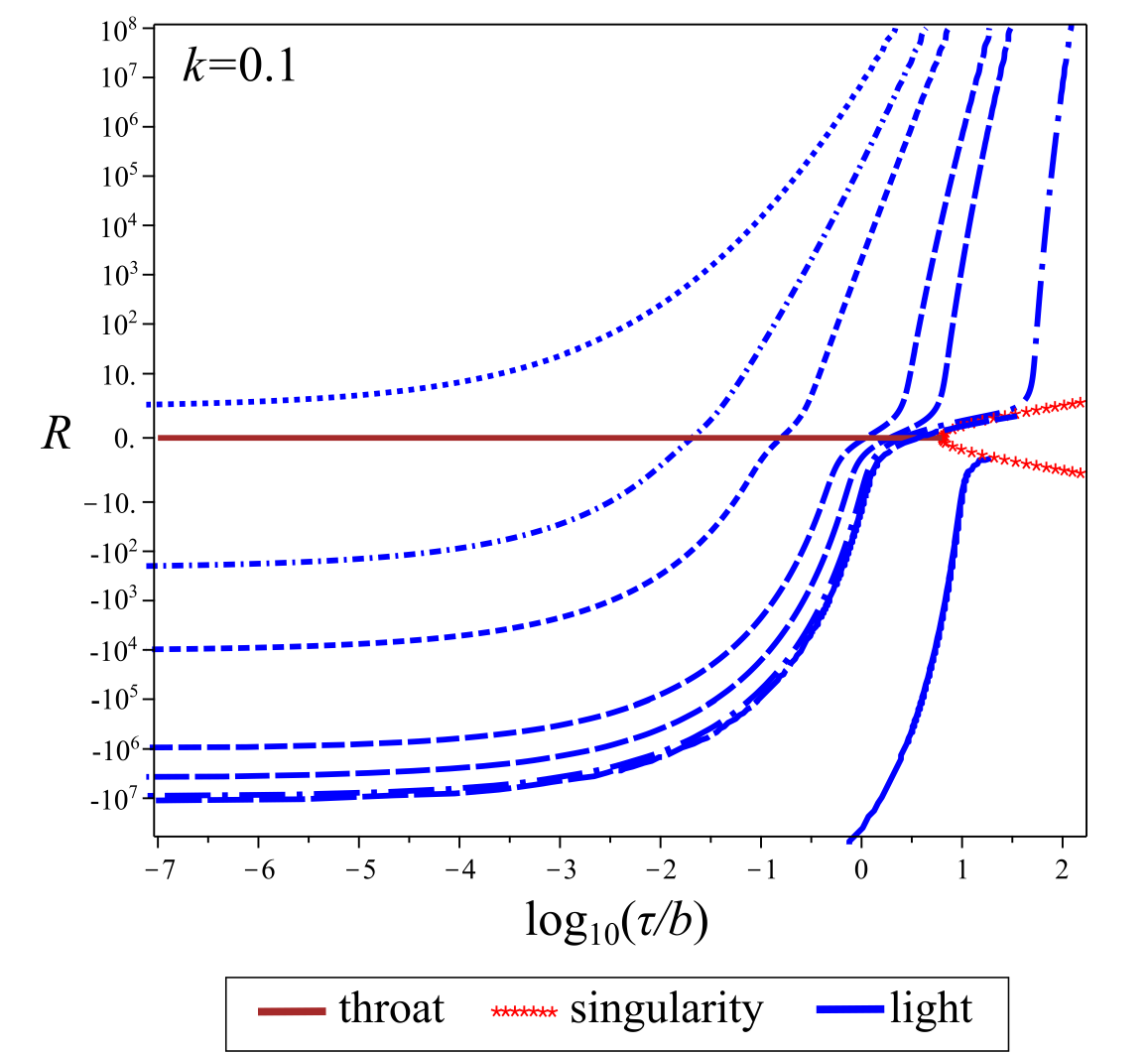} 
    \end{minipage} %\hspace{12pt}
\begin{minipage}{0.5\textwidth}
 \centering
\begin{tabular}{|c|c|c|}
\hline 
 $R_{*}$ & $b$ [cm] & $r_{*}$ [cm] \\ \hline\hline 
$10$ & $6.2\times 10^{24}$  & $2.0\times 10^{27}$ \\ \hline 
$10^2$ & $4.0\times 10^{21}$  & $2.0\times 10^{26}$ \\ \hline 
$10^3$ & $2.5\times 10^{18}$  & $2.0\times 10^{25}$ \\ \hline 
$10^4$ & $1.6\times 10^{15}$  & $2.0\times 10^{24}$ \\ \hline 
$10^5$ & $1.0\times 10^{12}$  & $2.0\times 10^{23}$ \\ \hline 
$10^6$ & $6.3\times 10^8$  & $2.0\times 10^{22}$ \\ \hline 
$10^7$ & $4.0\times 10^5$  & $2.0\times 10^{21}$ \\ \hline 
$10^8$ & $251$  & $2.0\times 10^{20}$  \\ \hline 
\end{tabular}
\end{minipage}
\caption{\small
	The figure shows the $\tau$-dependence of the radial coordinate $R$ of photons
	 at their radial motion (eight blue curves).  The brown 
	 horizontal line presents the throat $R=0$, the red line depicts the singularity $\eta=2\pi$. 
	 The table shows the correspondence between the junction coordinate $R=R^{*}$, the throat 
	 size $b$, and the radius $r_{*}$ of the \wh\ region. 
{The curves from top to bottom correspond to the following initial data at the moment 
	$\tau/b=10^{-7}$: $R=2.3$, $-198$, $-9.7\times 10^3$, $-9.6\times 10^5$, 
	$-3.7\times 10^6$, $-1.0\times 10^7$, $-1.1\times 10^7$, $-1.6\times 10^{11}$. }
	 }
	 \label{photon_01}
\end{figure}
% ----------------------------------------------------

\begin{figure}[h]
 \centering
\includegraphics[scale=0.35]{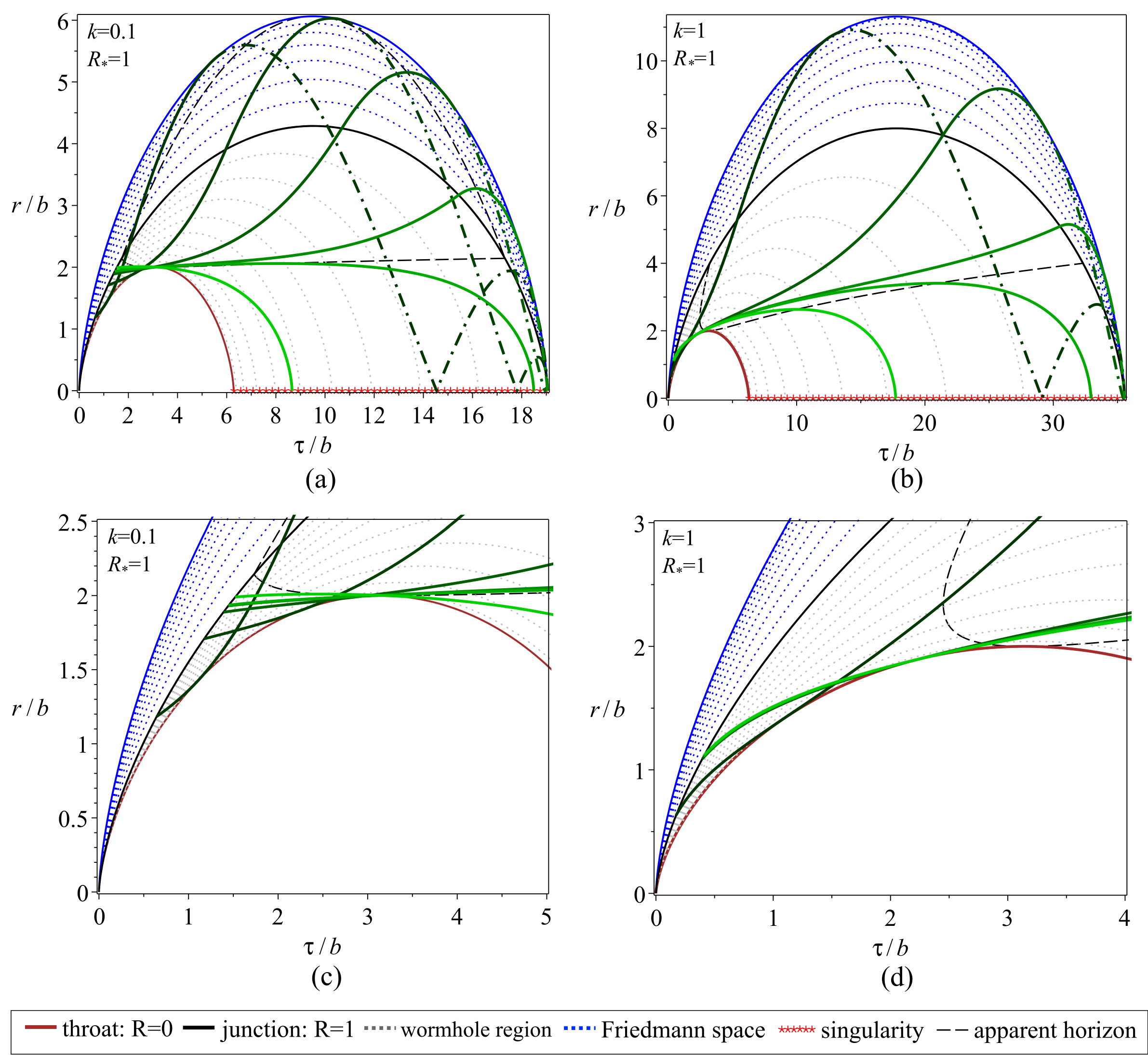} \\
\caption{\small
  	Illustrated are the dynamics of the throat $R=0$ (brown curve), the junction surfaces 
  	$R=\pm R_*$ (black curve), dust layers of the internal wormhole region $|R|\leqslant R_*$ 
  	(black point curves) and the external Friedmann universe $\chi\geqslant\chi_*$ (blue curves) 
  	in the case $R_*=1$.  Green curves correspond to photons moving from the region $R<0$ 
  	to $R>0$. The left figures (a) and (c) corresponds to the model with $k=0.1$, the right ones 
  	(b) and (d) to that with $k=1$. The results are presented in two scales. The top figures (a), 
  	(b) correspond to the usual scale, while the bottom figures (c), (d) correspond to an 
  	enlarged scale and clarify the dynamics at early times. The photons start their motion on 
  	the sphere $R=-R_*$, pass through the throat, and some of them leave the wormhole 
  	region in a finite time and move further in the Friedmann space-time. The dashed-dotted 
  	green curves correspond to the motion in the region $\chi\in(\pi/2;\pi)$ of the Friedmann 
  	universe. The red line presents the singularity $r=0$, and the black dashed curve 
  	corresponds to the apparent horizon. \label{figure2}}
\end{figure}

\begin{figure}[h]
\centering
\includegraphics[scale=0.35]{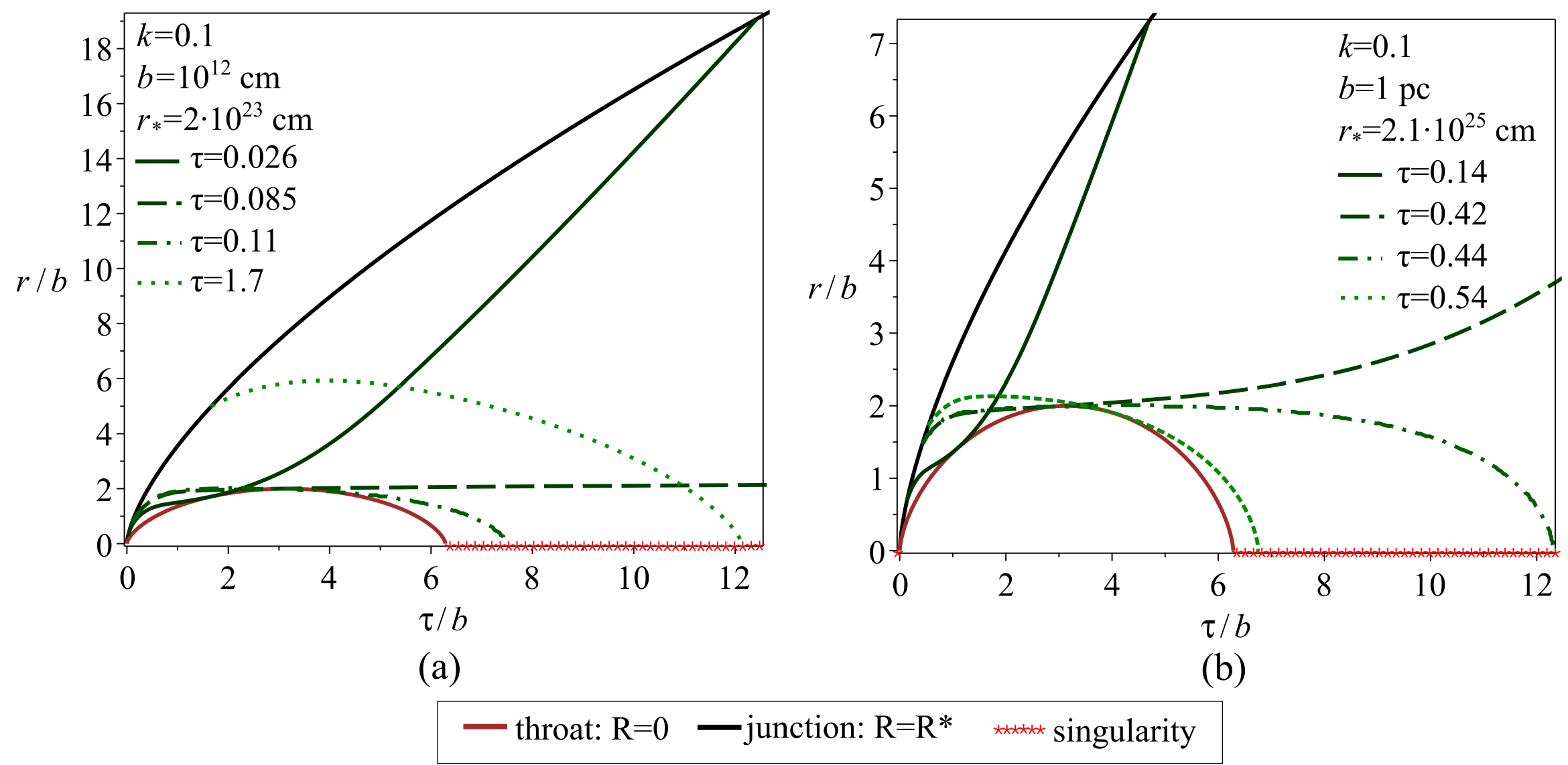}  \\
\caption{\small
	Dynamics of the wormhole throat $R=0$ (brown curve), the junction surface $R=\pm R_*$ 
	(black curve), photon trajectories (green curves) in the cases:
	(a) $k=0.1$, $b=10^{12}$ cm, $R_*=10^5$, $r_*=2\times 10^{23}$ cm; 
	(b) $k=0.1$, $b=1$ pc, $R_*=937$, $r_*=2.1\times 10^{25}$ cm. 
	The photons are launched on the surface $R=-R_*$ at different times: 
	(a) $\tau/b=0.026$, 0.085, 0.11 and 1.7; 
	(b) $\tau/b=0.14$, 0.42, 0.44, 0.54. One of the curves does not reach the throat, the rest ones 
	pass through the throat, and two of them reach the junction surface $R=R_*$. \label{figure3}}
\end{figure}

  The results of numerical integration of \eq (\ref{eq_ph}) or (\ref{dr_dtau}) are shown in 
  Figs.\,\ref{photon_01}, \ref{figure2}, \ref{figure3} and \ref{figure5}.  
   Figure \ref{photon_01} shows the set of null radial geodesics (blue curves) in the case $k=0.1$, 
  presented in the coordinates $(R,\tau)$. Graph uses a log-10 scale for just the $\tau$ axis.  
  The red line in the figure presents the singularity $\eta=2\pi$. There is also a singularity at the 
  initial time $\tau\to0$ ($\eta\to 0$), it is not presented. The photons begin their motion at the
  time instant with $\tau/b=10^{-7}$, close to the origin of the universe, and move from the 
  region $R<0$ to $R>0$ through the 
  wormhole region. The throat $R=0$ is shown in brown and exists 
  for a short time as compared to the universe lifetime. Some of the photons pass through the throat, 
  others fall to the singularity instead of reaching the throat. Further on, this solution must be glued 
  at some $R=R_*$ to the external Friedmann space-time, and the table on the right shows 
  the correspondence between the parameter $R_*$ and the radius $r_*$.

% --------------------------------------------------- fig 4
% -------------------------------------------

  Of greatest inteest are large values of the parameter $R_*$ (see Tables \ref{table1}
  and \ref{table2} above), however, for small enough $R_*=1$ ($\chi_*=\pi/4$, $a_0=4\sqrt2 b$) 
  the results are qualitatively similar and more suitable for illustration. Figure \ref{figure2} shows 
  the dynamics of the throat $R=0$ (brown curve),  
  dust layers of the wormhole region $|R|\leqslant R_*$ (black point curves), the junction surfaces 
  $R=\pm R_*$ (black curve), external Friedmann space-time $\chi\geqslant\chi_*$ (blue curve) 
  in the case $R_*=1$, presented in the coordinates $(\tau/b,r/b)$. The left panels (a) and (c)
  correspond to the model with $k=0.1$, the right ones (b) and (d) to that with $k=1$. The results 
  are presented in two scales: panels (a), (b) correspond to the usual scale, (c) and (d) to an 
  enlarged scale and clarify the dynamics at early times.The green curves correspond to 
  photons launched on the sphere $R=-R_*$ and moving from the region $R<0$ to $R>0$. 
  The geodesics in the left panels (a) and (c) are results of numerical integration in the 
  case $k=0.1$ with the following initial data at $R=-1$: $\tau/b=0.64$, $1.17$, $1.39$, $1.44$, 
  $1.45$, $1.52$. The right panels (b) and (d) present geodesics with the following initial data 
  at $R=-1$: $\tau/b=0.172$, $0.395$, $0.406$, $0.407$, $0.414$. The matter layers begin 
  and end their evolution at the singularity (red line). The purple curve corresponds to the 
  apparent horizon in the region $R>0$. 

  Note that to describe the motion in the $(\tau,r)$ coordinates, in fact, a set of two diagrams 
  is required, but due to their identity only one of them is shown. Each diagram in the figures 
  actually depicts two identical space-time regions $R\leqslant 0$ and $R\geqslant 0$, 
  connected by the throat $R=0$. 

% -------------------------------------------------- fig 5
% ---------------------------------------------------

  Figures \ref{figure3}a and \ref{figure3}b show the time dependence of the radius 
  $r(\tau,R)/b$ for photon paths (green curves) with the following parameter values: 
	(a) $k=0.1$, $b=10^{12}$~cm, $R_*=10^5$, $r_*=2\times 10^{23}$~cm, and 
	(b) $k=0.1$, $b=1$~pc, $R_*=937$, $r_*=2\times 10^{23}$~cm, 
  respectively. Unlike the previous figure, these values $b$ correspond to a realistic scale of 
  the model (see Tables \ref{table1} and \ref{table2}). 
  The photons are launched on the junction surface $R=-R_*$ at different times and move from 
  $R<0$ to $R>0$. Not all photons cross the throat and get to $R>0$, and some of them 
  reach the junction surface $R = R_*$ in finite time and enter the outer space-time. 
  In the left panel, the photons start from $R=-R_*$ with the initial data 
  $\tau/b=0.026$, $0.085$, $0.11$, $1.7$. In the right panel, the photons start from $R=-R_*$
   at the times $\tau/b=0.14$, $0.42$, $0.44$, $0.54$. 
  For example, the value $\tau/b=0.42$~cm corresponds to the time $t=\tau b/c=1.4$~year.
   We can conclude that the wormhole is traversable at least during a short time of its evolution. 

% -----------------------------------------------------------------------------------
\subsection{Radial motion of photons in the case $\tau_0\neq 0$}
% -----------------------------------------------------------------------------------

  Now let us consider radial null geodesics in the case (\ref{ex1-r}), 
  where $\tau_0(R)$ is a nonzero even function of $R$. The meaning of $\tau_0(R)$ is the time 
  $\tau$ (by the clock of an observer in Friedmann space-time) at which the dust layer 
  corresponding to a value of the $R$ coordinate begins to evolve. In particular, $\tau_0(0)$ is 
  the instant at which emerges the \wh\ throat $R=0$; this $\tau_0(0)$ can take arbitrary values 
  from the interval $0\leqslant\tau_0(0)\leqslant \tau_{\max}$,
  where $\tau_{\max}=\tau\big|_{\eta=2\pi,R=R_{*}}=2\pi b(1+R_*^{2})^{k+3/2}$ is the 
  lifetime of the universe. Different dust layers must not collide, therefore we must have $r' \ne 0$
  everywhere outside the throat. This condition is sufficient for the absence of a singularity, so that 
  the density (\ref{rho}) and the Kretschmann scalar (\ref{Kre}) are finite. Let the function 
  $\tau_0(R)$ vanish at the boundary $R=R_*$ of the \wh\ region, so that it does
  not affect the junction conditions. 

  The inequality $r' \ne 0$ is satisfied if we consider the following example of the function 
  $\tau_0(R)$:
\beq 			\label{tau0}
      \tau_0(R)= A \left[(R_*^{2}+1)^{k+3/2}-(R^{2}+1)^{k+3/2}\right],\qq 
      0\leqslant A\leqslant 2\pi b.
\eeq
  The derivative $r'$ of the function $r(R, \tau)$ has the form
\bear
	 r'= \frac{R(R^2+1)^k}{1-\cos\eta}\Big\{  A(2k+3)\sin\eta+
		b\left[4 k +5-\cos^{2}\eta-4(k +1) \cos \eta-(2k +3)\eta \sin\eta\right]  \Big\},
\ear
  where, as can be directly verified, the expression in curly brackets is positive, hence the condition 
  $r'\neq 0$ is satisfied at $R>0$ or $R<0$, and the density $\rho$ is everywhere finite and positive. 
  
% ---------------------------------- fig 6
\begin{figure}[h] 
\centering
 \includegraphics[scale=0.32]{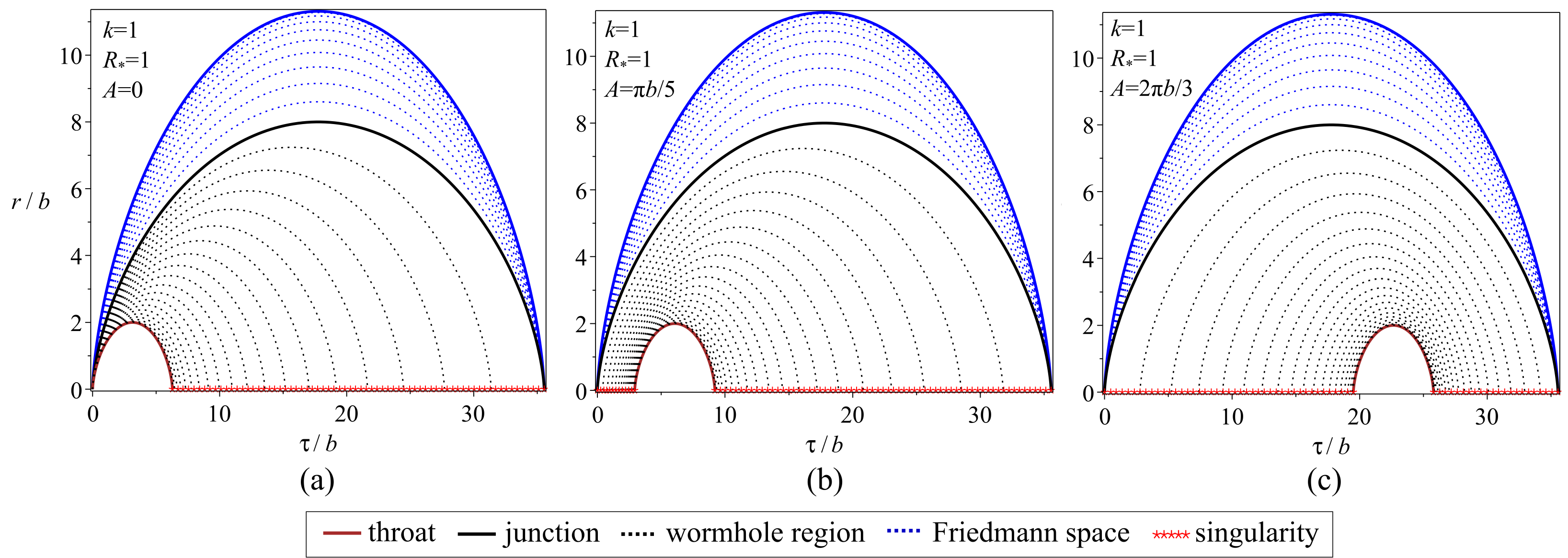}\\
\caption{\small
	The figure shows the dynamics of the dust layers in the case $k=1$, $R_*=1$ for different values of the parameters $A=0$, $\pi b/5$, $2\pi b/3$. \label{figure4}}
\end{figure}
% ------------------------------------------------

  In this model, the lifetime of the wormhole throat remains unchanged, equal to $2\pi b$,
  but different values of the parameter $A$ correspond to different emergence times of the 
  wormhole throat (Fig. \ref{figure4}). In 
  particular, in the case $A=0$ the throat begins to evolve simultaneously with all dust layers. 
  If we assume $A=2\pi b$, the throat collapses simultaneously 
  with all dust layers (such fine tuning looks quite incredible but still possible in principle). 
  Under the condition $0<A<2\pi b$, the throat emerges and collapses at intermediate times during 
  the lifetime of the universe. Smaller values of the parameter $k$ correspond to 
  a more compact wormhole region. Fig.\,\ref{figure4} corresponds to the case $R_*=1$ 
  ($a_0/b=4\sqrt2$), however, realistic values of the parameters $a_0$ and $b$ do not 
  change the qualitative picture of the system dynamics.

% ------------------------------------------ fig 7
\begin{figure}[h]
 \centering
\includegraphics[scale=0.4]{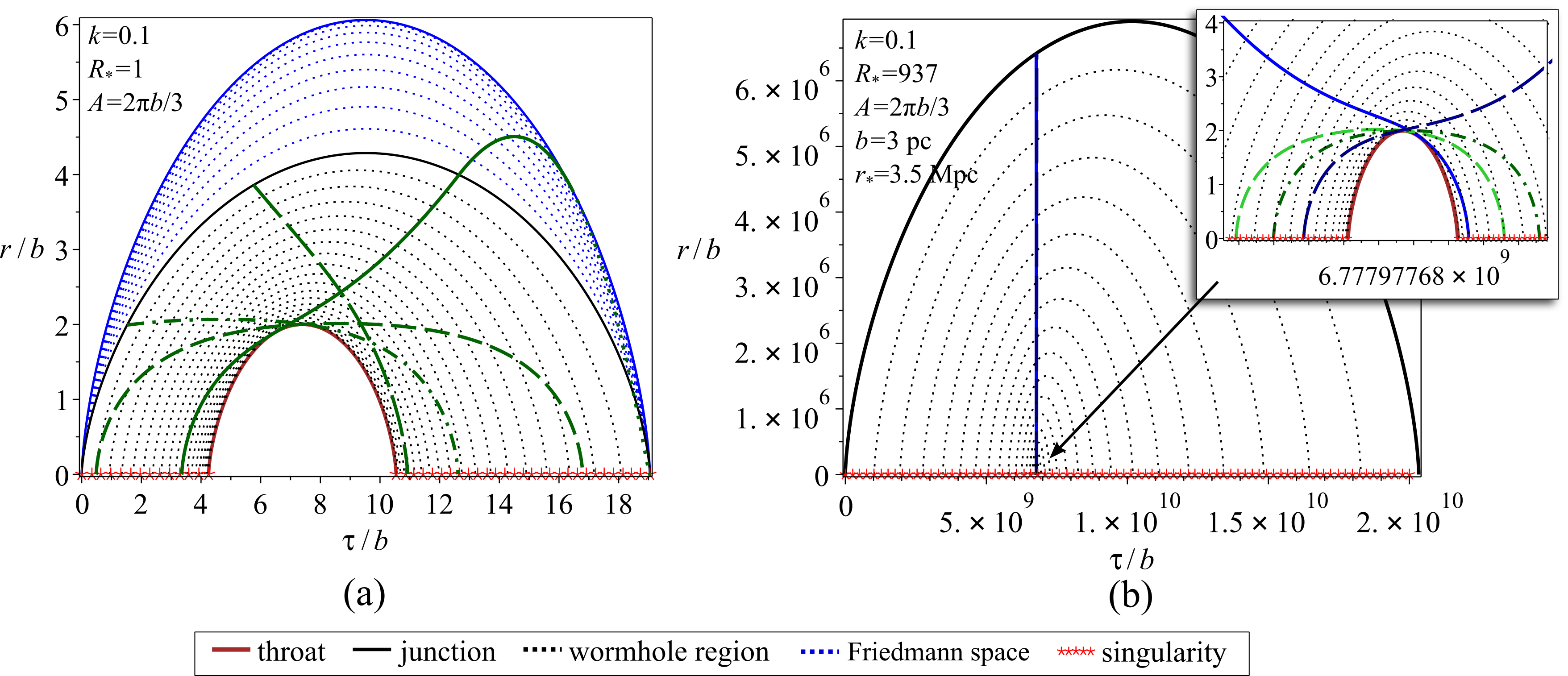} 
\caption{\small
	Dynamics of dust layers and radial photon trajectories. Left panel: $A=2\pi b/3$, $k=0.1$,
	$R_*=1$. Right panel: $A=2\pi b/3$, $k=0.1$, $R_*=937$, $b=3$ pc, $r_*=3.5$ Mpc.
 \label{figure5}}
\end{figure}
% --------------------------------------------------------------------
  Obviously, there are photons passing through the wormhole in the case of a thin dust region 
  $|R|\leqslant R_*$. However, as follows from the numerical estimates in Table 1, the case 
  of a thin dust region is of little interest. As noted in the section above, the model is traversable 
  with $\tau_0(R)=0$ ($A=0$). The value $A=2\pi b$ corresponds to the case where the 
  wormhole and the whole universe collapse simultaneously, it is quite similar to the case $A=0$ 
  and differs only by the direction of motion; in this case the wormhole region is always 
  traversable, at least for photons starting at times sufficiently close to the collapse time. 
  Due to continuity of the equations, the traversability is also expected for $A$ close enough 
  to zero or $2\pi b$. 
	
  The results of our numerical analysis are shown in Fig.\,\ref{figure5} for the case $k=0.1$, 
  $A=2\pi b/3$ in two versions. The throat emerges and collapses at some intermediate 
  times during the universe evolution since the parameter $A$ significantly differs from its 
  minimum ($A=0$) and maximum ($A=2\pi b$) values. Fig.\,\ref{figure5}a corresponds to 
  small enough $R_*=1$, in this case the results are qualitatively similar and more suitable 
  for illustration. 

  Figure\,\ref{figure5}b is obtained for values more consistent with cosmic scales, namely, 
  $b=3$\,pc, $r_*=3.5$\,Mpc, $R_*=937$. The inset in the right panel illustrates the
  behavior of the trajectories on a larger scale. In this case, there are no geodesics passing 
  through the whole wormhole area $|R|\leqslant R_*$. However, the throat is halfway 
  traversable, photons from the universe $R<0$ can get into the region $R>R_*$ if they 
  are emitted close enough to the throat. The trajectories are shown in blue for photons 
  passing through the throat and reaching $R=-R_*$ or $R=R_*$; the green color shows 
  trajectories passing through the throat but not leaving the wormhole area.

  As a result of our numerical analysis, we can conclude the following. In the general case, 
  the wormhole region $|R|\leqslant R_*$ can be traversable, but only under a particular choice 
  of the throat parameters and initial conditions. For any value of the junction surface $R_*$, 
  there are always light rays passing through the wormhole, at least for $A$ close enought 
  to zero or $2\pi b$. If the throat emerges in the middle part of the universe lifetime, 
  photons from the universe $R<0$ can get into the region $R>R_*$ if emitted close enough 
  to the throat. 
	
% ------------------------------------------------------------  
\subsection{Multiple wormholes in a multi-universe}
% ------------------------------------------------------------

\begin{figure}[h]
	\centering
	\includegraphics[scale=0.8]{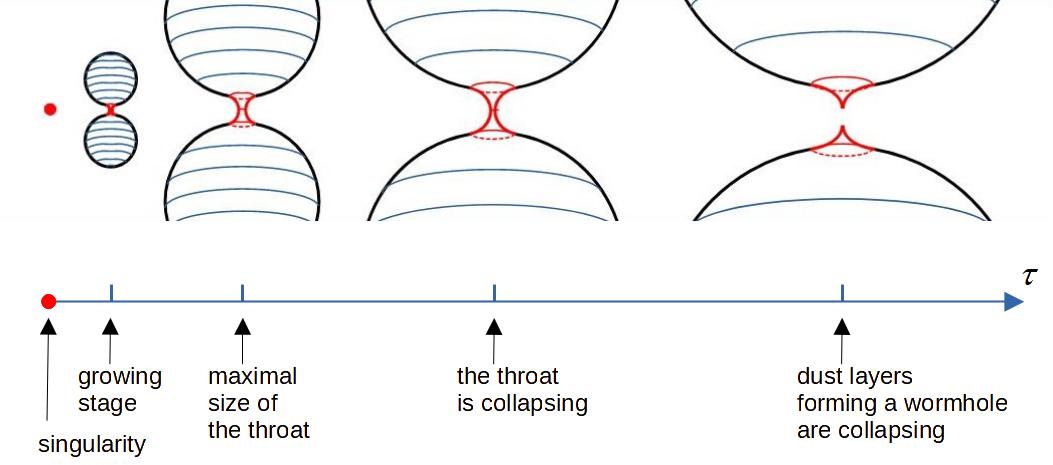}
	\caption{\small
		An evolving wormhole connecting two Friedmann universes}
	\label{evolvingwh}
\end{figure}

  Schematically, an evolving dust-filled configuration with a wormhole connecting two closed 
  Friedmann universes can be constructed as follows (see Fig. \ref{evolvingwh}). One takes two 
  copies of such universes, cuts off from each universe a three-dimensional spherical region, and 
  glues to the spherical boundaries being mouths of a dust-filled wormhole. This configuration 
  evolves synchronously with the proper cosmic time $\tau$, which is supposed to be the same in 
  all regions, from the initial cosmological singularity to the final one. It is worth noting that the 
  wormhole mouths inscribed into closed Friedmann universes are existing the entire 
  cosmological cycle and evolving synchronously with the universe evolution, i.e., growing at the 
  expansion phase and shrinking at contraction. On the other hand, the wormhole throat 
  situated between the two mouths is only open during a small interval of the universe evolution. 
  Figure \ref{evolvingwh} shows an example where a throat appears at the moment of initial 
  singularity, then it grows, achieves its maximum size, and after that shrinks and disappears. 
  In Fig.\,\ref{figure4}, one can see other examples where wormhole throats appear during the 
  cosmological evolution.   
% -------------------------------------------------------- fig 8
% -----------------------------------------------------------

\begin{figure}[h]
	\centering
	\includegraphics[scale=0.5]{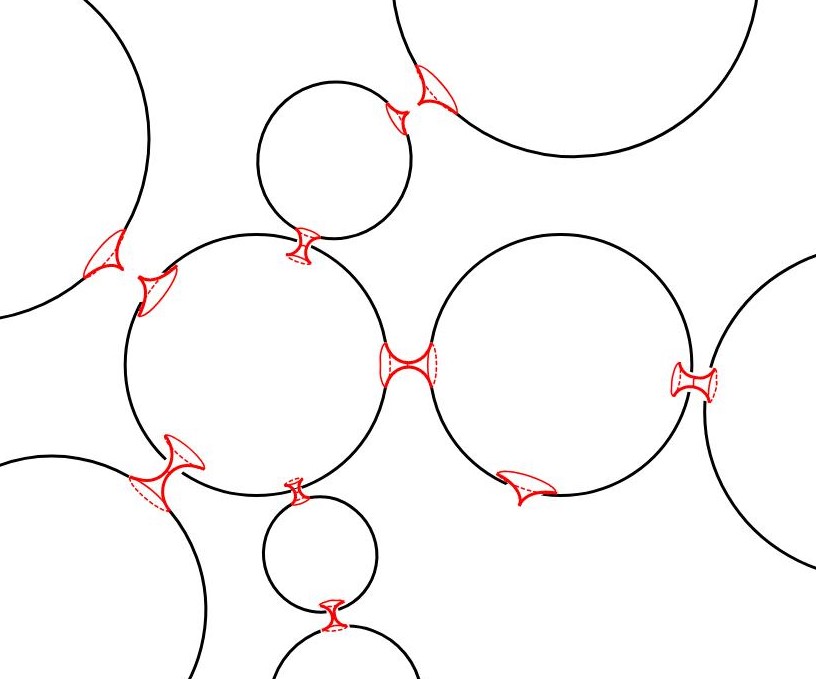}
	\caption{\small
		Multiple wormholes in the multi-universe
		\label{multiplywh}}  
\end{figure}

  The model with one dust-filled wormhole connecting two closed Friedmann universes can be 
  naturally generalized. We can suppose that the ``mother'' Friedmann universe is born with 
  multiple mouths of wormholes associated to ``daughter'' universes. As a result, we obtain a 
  model of a multi-universe as a system of closed Friedmann universes connected by evolving 
  dust-filled wormholes (see Fig. \ref{multiplywh}).

  Here it is necessary to stress once more that the multi-universe with multiple wormholes evolves 
  synchronously with  unified proper cosmic time $\tau$, which is supposed to be the same in all 
  regions. Such a high correlation between different regions can be explained if one supposes that 
  the multi-universe is born from the quantum spacetime foam on sub-Planckian scales as a single 
  quantum state.

  One more point worth being stressed is the following.
  Strictly speaking, the Friedmann universe with an inscribed wormhole mouth is already neighter 
  homogeneous nor isotropic. A distant observer will see a wormhole mouth as a compact object 
  bending photon trajectories. In addition, a wormhole mouth will introduce distortions into the 
  spectrum of the almost isotropic cosmic microwave background radiation. The scale of anisotropy 
  must be proportional to an angular size of the mouth. In principle, these both effects could be 
  potentially observable, therefore, one might verify the model of dust-filled wormholes in the 
  Friedmann universe using astrophysical methods. Particular predictions of this kind 
  require a further study. 
% ------------------------------------------------------------------- fig 9
% ---------------------------------------------------------------

% ========================
\section{Concluding remarks} 
% ========================
  
  In this paper, we have continued our study begun in \cite{we-wh21} and described   
  in some detail different features of evolving \whs\ able to exist in a Friedmann universe 
  in the simplest case of purely dust solutions. However, it is evident that adding small values of 
  the cosmological constant $\Lambda$ cannot qualitatively change such local issues
  as the existence and properties of wormholes. Meanwhile, a nonzero $\Lambda$ drastically 
  changes the global dynamics: $\Lambda > 0$ launches a stage of accelerated expansion of the 
  Universe, which must probably encompass the wormhole region. It is important that such \wh\ 
  regions can exist not only at a matter-dominated stage of the Universe evolution but also at its
  accelerated stage. In particular, examples of solutions to the Einstein equations describing \whs\ 
  in a de Sitter universe are known, and it has been noticed that such \whs, if they existed at an 
  inflationary stage, could greatly extend the causal connection of different parts of the universe 
  \cite{kb17}.
  
\begin{figure}[h]
 \centering
\includegraphics[scale=0.6]{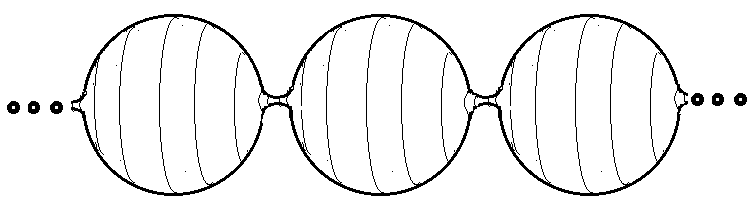} 
\caption{\small
		Multiple universes connected by magnetic wormholes}
 \label{figure10}
\end{figure}

  On the other hand, inclusion of a sufficiently small charge $q\ne 0$ also cannot strongly change the 
  local picture of a wormhole. However, globally, the Universe cannot be precisely homogeneous and 
  isotropic in the presence of a vector field. Also, a charge (or an effective charge due to a \wh) on 
  one ``pole'' inevitably leads to an opposite charge on the other, where the lines of force again 
  converge. There can be a similar wormhole mouth at this other pole and one more universe 
  beyond it, and so on. The whole picture will resemble a ``churchkhela,'' wonderful Georgian 
  dessert, see Fig.\,\ref{figure10}. As before, here can also be natural generalizations in the spirit
  of Fig.\,\ref{multiplywh}, not to mention that some of the \whs\ may connect different parts
  of the same universe. Possible observational signatures of such objects, in particular, concerning the 
  properties of cosmic microwave background and cosmic magnetic fields, can be a subject of further 
  studies. 
  
% ------------------------------------------ fig 10
% --------------------------------------------------------------------  
   
% ================================      
\Acknow{P.E.K. and S.V.S. are supported by RSF grant No. 21-12-00130. 
 Partially, this work was done in the ramework of the Russian Government Program of 
 Competitive Growth of the Kazan Federal University.
 K.B. was supported in part by the RUDN Project No. FSSF-2023-0003, and by the 
 Ministry of Science and Higher Education of the Russian Federation Project 
 ``Fundamental properties of elementary particles and cosmology'' N 0723-2020-0041.}
  
% ================================ 

\end{document}